\documentclass[manuscript,screen]{acmart}

\AtBeginDocument{%
  \providecommand\BibTeX{{%
    \normalfont B\kern-0.5em{\scshape i\kern-0.25em b}\kern-0.8em\TeX}}}

\setcopyright{acmcopyright}
\copyrightyear{2022}
\acmYear{2022}
\acmDOI{XXXXXXX.XXXXXXX}

\acmPrice{15.00}
\acmISBN{978-1-4503-XXXX-X/18/06}

\usepackage{soul}
\usepackage{hyperref}
\usepackage{multirow}
\usepackage{listings}
\usepackage{fancybox}
\usepackage{graphicx}
\usepackage{color}
\usepackage{xcolor}
\usepackage{array}
\usepackage{amsmath}
\usepackage{xspace}
\usepackage{subcaption}
\usepackage{url}
\usepackage{tikz}
\usepackage{caption}
\usepackage{pgfplots}
\pgfplotsset{compat=1.16}
\usepackage{pgf-pie}
\usetikzlibrary{pgfplots.statistics,calc}
\usepackage{algorithm}
\usepackage{algorithmic}

\usepackage{tcolorbox}
\makeatletter
\newcommand{\mybox}[1]{%
	\setbox0=\hbox{#1}%
	\setlength{\@tempdima}{\dimexpr\wd0+13pt}%
	\begin{tcolorbox}[boxrule=0.5pt, colback=white, arc=4pt,
		left=6pt,right=6pt,top=6pt,bottom=6pt,boxsep=0pt]
		#1
	\end{tcolorbox}
}

\definecolor{songcolor}{RGB}{191,191,191}

\begin{document}

\title{The Good, the Bad, and the Missing: Neural Code Generation for Machine Learning Tasks}

\author{Jiho Shin}
\email{jihoshin@yorku.ca}
\orcid{0000-0001-8829-3773}
\affiliation{%
  \institution{York University}
  \streetaddress{4700 Keele St.}
  \city{North York}
  \state{Ontario}
  \country{Canada}
  \postcode{M3J 1P3}
}

\author{Moshi Wei}
\affiliation{%
  \institution{York University}
  \streetaddress{4700 Keele St.}
  \city{North York}
  \country{Canada}}
\email{moshiwei@yorku.ca}

\author{Junjie Wang}
\affiliation{%
  \institution{Institute of Software, Chinese Academy of Sciences}
  \city{Beijing}
  \country{China}}
  \email{junjie@iscas.ac.cn}

\author{Lin Shi}
\affiliation{%
  \institution{Institute of Software, Chinese Academy of Sciences}
  \city{Beijing}
  \country{China}}
  \email{shilin@iscas.ac.cn}

\author{Song Wang}
\affiliation{%
  \institution{York University}
  \streetaddress{4700 Keele St.}
  \city{North York}
  \country{Canada}}
\email{wangsong@yorku.ca}

\renewcommand{\shortauthors}{Shin et al.}

\begin{abstract}

Machine learning (ML) has been increasingly used in a variety of domains, while solving ML programming tasks poses unique challenges because of the fundamentally different nature and construction from general programming tasks, especially for developers who do not have ML backgrounds. 
Automatic code generation that produces a code snippet from a natural language description can be a promising technique to accelerate ML  programming tasks.
In recent years, although many deep learning-based neural code generation models have been proposed with high accuracy, the fact that most of them are mainly evaluated on general programming tasks calls into question their effectiveness and usefulness in ML programming tasks.
In this paper, we set out to investigate the effectiveness of existing neural code generation models on ML programming tasks.
For our analysis, we select six state-of-the-art neural code generation models, and evaluate their performance on four widely used ML libraries, with newly-created 83K pairs of natural-language described ML programming tasks.
Our empirical study reveals some good, bad, and missing aspects of neural code generation models on ML tasks, with a few major ones listed below. 
(\textbf{Good}) Neural code generation models perform significantly better on ML tasks than on non-ML tasks.
(\textbf{Bad}) Most of the generated code is semantically incorrect.
(\textbf{Bad}) Code generation models cannot significantly improve developers’ completion time.
(\textbf{Good}) The generated code can help developers write more correct code by providing developers with clues for using correct APIs.
(\textbf{Missing}) The observation from our user study reveals the missing aspects of code generation for ML tasks, e.g., decomposing code generation for divide-and-conquer into two tasks: API sequence identification and API usage generation.

\end{abstract}



\received{20 February 2022}
\received[revised]{12 March 2022}
\received[accepted]{5 June 2022}

\maketitle

\section{Introduction}
\label{sec:intro}
Recently, with the advances in deep learning techniques~\cite{vaswani2017attention,devlin2018bert}, 
many neural code generation models have been proposed and extensively studied~\cite{dong2016language,yin2018tranx, sun2020treegen,guo2020graphcodebert,norouzi2021code}.
Most of these models treat code generation as a machine translation task and are built in an end-to-end manner, i.e., a corpus of source code and comment pairs are used to train the models, and given a natural language (NL) described programming task, the models will generate a source code snippet.  

Although many neural code generation models have exerted high accuracy in generating source code from NL described programming tasks, most of which are mainly evaluated on general programming tasks, little is known about their effectiveness and usefulness for domain-specific tasks, In this study, we investigate Machine Learning (ML), which has significantly different paradigms from traditional general programming tasks (relatively more deterministic and less statistically orientated) \cite{zhang2020machine}. 
In addition, ML programming tasks often 
require many complex algorithms, mathematical operations, and data process operations~\cite{nguyen2019machine,wang2021automatic}.

To investigate the effectiveness of existing neural code generation models on ML programming tasks, We use six state-of-the-art 
neural code generation models as the baselines (details are in Section~\ref{sec:state-of-the-art}) and we select four widely used ML libraries, i.e., Scikit-Learn\footnote{\url{https://scikit-learn.org/stable/}}, Keras\footnote{\url{https://keras.io/}}, TensorFlow\footnote{\url{https://www.tensorflow.org/}}, and PyTorch\footnote{\url{https://pytorch.org/}} for ML programming task collection. 
In addition, inspired by existing studies~\cite{wang2021automatic}, we {categorize the ML programming tasks according to the utilized APIs into three different categories based on their purposes in the ML pipeline}, i.e., \emph{data process}, \emph{model building}, and \emph{evaluation} (details are in Section~\ref{sec:mltasks}).  


For our analysis, we need a new dataset of ML programming tasks inclusively as most of the existing widely used code generation benchmark datasets, e.g., StaQC~\cite{yao2018staqc}, CoNaLa~\cite{yin2018learning}, Django~\cite{oda2015learning}, and ReCa~\cite{liu2020deep} mainly contain general-purpose programming tasks. Specifically, we create the new ML task-related dataset by reusing data from JuICe~\cite{agashe2019juice}, i.e., a refined human-curated code generation dataset including 1.5 million examples mined from over 659K publicly available Jupyter notebooks from GitHub. 
We use the API information from the four studied ML libraries to extract ML-related programming tasks.
Note that, we also have {collected} a non-ML task-related dataset from JuICe~\cite{agashe2019juice} as a comparison dataset.
We then evaluate the six selected state-of-the-art neural code generation models on the two datasets (ML vs non-ML) regarding the accuracy, syntactic and semantic correctness, and usefulness of the generated code. 
Our empirical study reveals some good, bad, and missing aspects of state-of-the-art neural code
generation models on ML tasks, with a few major ones listed below.
(\textbf{Good}) Neural code generation models perform significantly better on ML tasks than on non-ML tasks.
(\textbf{Good}) The generated code can help developers write more correct code by providing developers with clues for using correct APIs. 
(\textbf{Bad}) Most of the generated code is semantically incorrect. 
(\textbf{Bad}) Code generation models cannot significantly improve developers’ completion time.
(\textbf{Missing}) The observation from our user study reveals the missing aspects of code generation for ML tasks, e.g., decomposing code generation for divide-and-conquer into two tasks: API sequence identification and API usage generation.



This paper makes the following contributions:

\begin{itemize}

\item The first empirical study for evaluating six state-of-the-art neural code generation models on ML programming tasks regarding the accuracy, syntactic and semantic correctness, and usefulness of the generated code. 

\item The missing viewpoints between existing automatic neural code generation models and practical ML programming tasks, as well as future research directions for improving code generation.  

\item A new dataset that is composed of 83K pairs of NL descriptions and the corresponding source code that are implemented by four ML libraries, which facilitate the follow-up studies in  this direction.

\item The released source code of our tool and the dataset of our experiments to help other researchers replicate and extend our study\footnote{\url{https://zenodo.org/record/7036255}}.
\end{itemize}

The rest of this paper is organized as follows. 
Section~\ref{sec:motivation} presents
the background of this study.  
Section~\ref{sec:approach} describes the methodology and the study design of our work. 
Section~\ref{sec:result} presents the evaluation results. 
Section~\ref{sec:discussion} discusses open questions and the threats to the validity of this work. 
Section~\ref{sec:related} presents the related studies. 
Section~\ref{sec:conclusion} concludes this paper. 
\section{Background}
\label{sec:motivation}
This section introduces the background of this study, i.e., the neural source code generation model and ML programming tasks.

\subsection{Neural Code Generation}
\label{sec:workflow}
Neural code generation techniques exploit deep neural generation models to learn the distribution or the density estimation of source code with different representations such as the sequence of tokens or tree structures such as abstract syntax tree (AST) \cite{dahal2021analysis,lyu2021embedding}. 
Due to the similarity of the data representation, the source code generation model inherit many of the techniques from NL generation models.
In NL generation, many tasks can be categorized such as content determination, text structuring sentence aggregation, lexicalization, referring expression generation, linguistic realization, and so on \cite{gatt2018survey}.
There are also different tasks in the source code generation such as code completion \cite{liu2016neural}, code search \cite{gu2018deep}, program synthesis via rules or examples \cite{cozzie2012macho}, domain-specific \cite{dong2018coarse} or general-purpose code generation \cite{bui2021self}, code to code translation \cite{ahmad2021avatar}, API recommendation \cite{chen2021holistic}, and so on.
The generation model that this paper focuses on is the multi(bi) modal modeling of source code and NL, which usually uses an end-to-end encoder-decoder architecture of the model that is widely used in neural machine translations. 
The encoder takes in the sequence of an NL description of a code and maps them in a latent space and then the decoder decodes them into a source code.
The encoder-decoder layers learn to map similar data points of each modal to be closely mapped in the latent space so that it generates similar source codes.

To evaluate code generation models, many different benchmark datasets have been proposed. Django~\cite{oda2015learning} dataset was first mined to generate pseudo-code from source code for the Django framework (Python-to-English). The data contains 18K pairs of python statements and their corresponding English pseudo-code.
CoNaLa~\cite{yin2018learning} is a dataset mined from Stack Overflow posts which consists of 3K pairs of human-annotated python code snippets and its NL intent. 
StaQC~\cite{yao2018staqc} is another source code and NL description pair that are mined from Stack Overflow. It consists of 1.4K Python and 1.2K SQL queries which were mined using a neural network to incorporate textual similarities. 
ReCa~\cite{liu2020deep} is a large-scale dataset of NL requirements and its programming language. ReCa consists of data for multiple general {languages}, i.e., C, C++, Java, and Python. It also has multiple programs for the same requirement.
JuIce~\cite{agashe2019juice} is a new large-scale open-domain dataset composed of 1.5M pairs of Python code examples and the corresponding NL descriptions collected from 659K publicly available Jupyter notebooks on GitHub.
\begin{table}[t!]
\centering
\caption{Statistics of existing benchmark datasets. \textbf{PL} denotes the program languages, \textbf{Loc} denotes the average lines of code, and \textbf{Length} denotes the average length of NL tasks in token.}
\label{tab:dataintro}
\begin{tabular}{|l|l|l|c|c|c|}
\hline
\textbf{Dataset} & \textbf{Domain} & \textbf{PL} & \textbf{\#Pairs} & \textbf{ Loc} & \textbf{Length} \\ \hline
Django&general   & Python &18.8K& 1 & 14 \\ \hline
CoNaLa&  general   &Python &2.9K& 1.1 &14 \\ \hline
StaQC&  general  &Python/SQL &2.6k& 10.1 &10 \\ \hline
ReCa &  general  &C/Java/Python &101K& 37.2& 185 \\ \hline
JuIce&  general  & Python&1.5M& 10.0 & 39.7\\ \hline
\end{tabular}
\end{table}
Table~\ref{tab:dataintro} shows more detailed statistics of these existing datasets.


%

\subsection{Machine Learning Tasks}
\label{sec:mltasks}
Machine learning-based tasks often contain different steps, which are also known as the pipeline of machine learning~\cite{amershi2019software}. 

Specifically, constructing an ML pipeline contains the following three steps.  
The first step is mainly about the processing of the dataset that is needed to train a model.
It involves pre-processing tasks such as data loading, noise filtering, format converting, data imputation, scaling or normalization, and feature engineering (e.g., feature selection, elimination, word embedding)~\cite{khalid2014survey}.
After processing the data, one needs to further specify the learning algorithm to infer relations that capture the hidden patterns in the dataset. 
In the specified algorithm, there is information regarding the parameters that are learned and fit the input data, the loss function to be minimized, the optimizer and the learning rate to lead the model in better learning, and the architecture of the model that specifies the construction of the network's layers and nodes~\cite{lecun2015deep}.
And finally, one needs to evaluate the trained model. This step includes choosing the evaluation metrics to assess the performance of trained models, e.g., accuracy, precision, and recall~\cite{huang2005using}.
Evaluations are applied to an unseen dataset that is separate from the training data to assess the generalizability of the trained model.
In this work, to understand the effectiveness of existing neural code generation models on different types of ML programming tasks, we categorize ML programming tasks into different steps in the ML pipeline~\cite{wang2021automatic}.


\begin{itemize}
    \item \textbf{\emph{Data processing}}:
    tasks that are related to data processing, e.g., data loading, format transformation, normalization,
    feature engineering, and data transformation.
    \item \textbf{\emph{Model building}}: 
    tasks that are related to building the
    machine/deep learning models, which can be conventional ML models such as classification models, regression models, clustering models, or neural network models with
    specific neural architecture regarding the type of layers and activation function used. 
    \item \textbf{\emph{Evaluation}}: 
    tasks that are related to evaluating the performance of the trained machine/deep learning models, e.g., getting the values of different evaluation metrics such as accuracy, f-measure, AUC, or getting the probability of instances from the model.
\end{itemize}
Note that while we focus on three types of ML programming tasks in the ML pipeline (i.e., \emph{data process}, \emph{model building}, and \emph{evaluation}), there are other tasks in the ML pipeline, e.g., model deployment and performance monitoring~\cite{amershi2019software}. 
As finishing these tasks mainly requires auxiliary functions or scripts rather than the APIs from the ML libraries, thus we exclude these tasks in this study.
\section{Study Design}
\label{sec:approach}

In this section, we discuss the design of our empirical study on evaluating the performance of code generation models on ML programming tasks. 
\begin{table}[t!]
\begin{center}
\caption{The studied ML libraries in this work.}
\begin{tabular}{|l|l|c|}
  \hline			
    \textbf{Library} & \textbf{Description} & \textbf{\#API} \\\hline
    Scikit-learn & A library for ML algorithms. & 1.2k\\\hline
    Keras & An interface for artificial neural networks. & 1.4k\\\hline
    TensorFlow &A library for DL developed by Google. & 8.0k\\\hline
    PyTorch & A library for ML and DL tasks. &0.4k \\\hline

\end{tabular}
\label{tab:study_subject}
\end{center}
\end{table}


\begin{table}[t]
\centering
\caption{Statistics of the experimental datasets. \textbf{\#API} is the average number of APIs involved. 
\label{tab:experimentData}
}
\begin{tabular}{|l|l|l|l|l|l|}
\hline
\textbf{ML task} & \textbf{\#Training} & \textbf{\#Testing} & \textbf{Loc} & \textbf{Length} &\textbf{\#API} \\ \hline
data & 30.9K & 5.5K & 3.93 & 12.67 & 1.48 \\ \hline
model & 26.3K & 5.5K & 4.18 & 12.44 & 1.28 \\ \hline
evaluation & 13K & 2.6K & 3.28 & 11.59 & 1.13 \\ \hline \hline
non-ml & 34K & 5.5K & 3.23 & 13.26 & 1.63 \\ \hline 
\end{tabular}
\end{table}
\begin{figure}[t!]
    \centering
      \includegraphics[width=0.8\columnwidth]{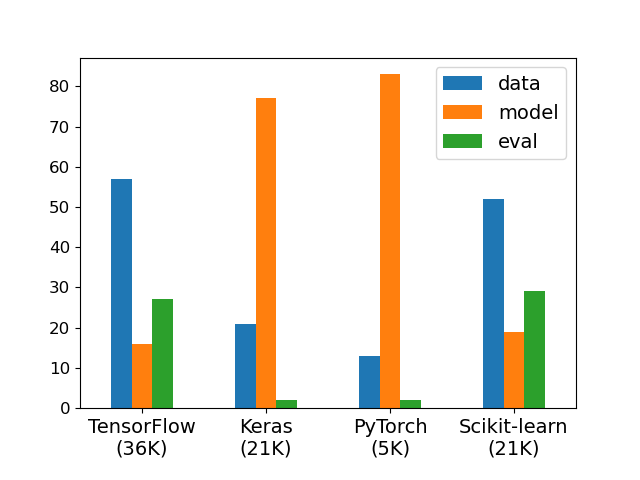}
      \vspace{-0.1in}
    \caption{Distribution of experiment data in each ML library. 
    }
   \label{fig:disc}
\end{figure}

\subsection{Data Collection}
\label{sec:data}


In this work, we prepare a new dataset that contains the three types of ML programming tasks (i.e., \emph{data process}, \emph{model building}, and \emph{evaluation}) as most of the existing widely used code generation benchmark datasets, e.g., StaQC~\cite{yao2018staqc}, CoNaLa~\cite{yin2018learning}, Django~\cite{oda2015learning}, and ReCa~\cite{liu2020deep} mainly contains general-purpose programming tasks. 
Specifically, we create our dataset from JuICe~\cite{agashe2019juice}, i.e., a refined human-curated code generation dataset including 1.5 million Python code examples mined from over 659K publicly available Jupyter notebooks from GitHub. 

\noindent \textbf{ML API selection}: The original JuICe dataset \cite{agashe2019juice} contains open-domain programming tasks. 
For our study, we focus on ML programming tasks that are developed by four widely used ML libraries, i.e., Scikit-learn, Keras, TensorFlow, and PyTorch. The list of studied libraries is organized in Table~\ref{tab:study_subject}.
Specifically, three of the authors independently use the following two steps to manually identify all the APIs in each of the four ML libraries that are related to the three types of ML programming tasks (i.e., \emph{data process}, \emph{model building}, and \emph{evaluation}) based on the documentation of these APIs: 1). We read the documentation of the API to understand its functionality. A category will be assigned to the API based on its major intent. 
2). If a decision cannot made based on the documentation of the API as the description might not contain enough information, we further check the source code following the instructions in Step 1 to label the API. The value of Fleiss' kappa during this process is 0.90, which shows an almost perfect agreement. {In total, 128, 244, and 51 APIs are identified for the three types of ML programming tasks, i.e., \emph{data process}, \emph{model building}, and \emph{evaluation} respectively.}

\noindent \textbf{ML-task data construction}: After we identify the APIs in an ML library for each of the three ML programming tasks, we further use the corresponding APIs as the filter to collect ML programming tasks from the JuICe dataset~\cite{agashe2019juice}, e.g., we scan the source code of each programming task in JuICe, a programming task will be labeled as \emph{data process} task if it only contains \emph{data process} related APIs from each of the four ML libraries.
We discard any instances that use multiple ML libraries {to distinguish programming tasks that use different libraries.}

\noindent \textbf{Data cleaning}: We also remove duplicated data instances,  meaningless source code elements, i.e.,  `$>>>$', single and multi-line comments, lines that start with `!' and `\%' for commands, code snippets with just imports, and none ASCII code. 
{We also apply additional filtering to further denoise the dataset and make the problem simpler by discarding instances with NL description lengths below 20 and over 150, and source code with more than 10 LOC as suggested by existing work~\cite{jozefowicz2016exploring}. To assess the quality of the collected data, the first three authors performed a manual analysis on 100 random samples from each splits, i.e. \emph{data processing}, \emph{evaluation}, and \emph{model building}. The authors first checked if the source code of the instance is relevant to the ML programming task. Then we check if the NL description is relevant to the source code. We found that 64\%-81\% of the samples have both source code and NL relevant to the ML programming task and its description. 
We also carefully split the instances on a project level, i.e. instances from the same project go to the same split, to avoid any information leakage.}


The new dataset of ML programming tasks is organized in Table~\ref{tab:experimentData}. 
We also show the distribution of the different ML programming tasks in each ML library in Figure~\ref{fig:disc}.

\subsection{Studied Neural Code Generation Models}
\label{sec:state-of-the-art}

In this work, we select six state-of-the-art neural code generation models proposed in the recent three years as our experiment subjects. These models are selected mainly based on the following inclusion criteria:
\begin{itemize}
    \item The model should not stick to a specific programming language, which makes it practical and generalizable for it to work on our dataset.
    \item The model's implementation should be publicly available and replicable, which can remove potential implementation bias and facilitates the evaluation process.
    \item The model should be proposed in recent years and represents the state-of-the-art, i.e., has a high performance in generation in BLEU score.
\end{itemize}

As a result, six neural code generation models are selected from 29 candidate papers published on ACL, ICLR, AAAI, TSE, EMNLP, etc.
Details of these models are as follows:

\vspace{4pt}
\noindent \textbf{tranX}~\cite{yin2018tranx}: proposed a transition-based neural semantic parser that maps NL utterances into formal meaning representations such as executable programs. Specifically, it develops a transition system that decomposes the generation procedure of an AST into a sequence of tree-constructing actions. 

\noindent \textbf{External-Knowledge-Codegen (EK Codegen)}~\cite{xu2020incorporating}: proposed a model-agnostic approach based on data augmentation, retrieval, and data re-sampling, to incorporate external knowledge into code generation models. Specifically, the mined NL-code pairs from the online programming QA forum Stack Overflow and API documentation for pre-training. The basic architecture of this model uses \textbf{tranX} as its base.

\noindent \textbf{CG-RL}~\cite{jiang2021exploring}: proposed an extended Seq2Tree model equipped with a context-based branch selector, which is capable of dynamically determining optimal branch expansion orders for multi-branch nodes. Specifically, it adopts reinforcement learning to train the whole model with an elaborate reward that measures the loss difference between different branch expansion orders.

\noindent \textbf{Codegen-TAE}~\cite{norouzi2021code}: exploited the use of a large amount of monolingual programming language corpus to fit a generic transformer-based Seq2Seq model with minimal code-generation-specific inductive bias.

\noindent \textbf{TreeCodeGen} \cite{dahal2021analysis}: proposed a Transformer-based structure-aware tree-to-tree model. They {adopted} a Tree Transformer, an attention-based tree-to-tree model with a hierarchical accumulation for code generation.

\noindent \textbf{PyCodeGPT}~\cite{zan2022cert}: proposed a large pre-trained language model based on the GPT-Neo~\cite{black2021gpt} with 110M parameters, which is comparable to the powerful GPT-3 model~\cite{brown2020language}.
They mined 1.2M Python repositories in GitHub and filtered/pre-processed high-quality Python code resulting in 96GB of training data.

In our experiment, to remove potential bias, we use the implementation of their publicly available code with the settings that {have} the best performance reported in their papers or used in their replication packages.

\subsection{Research Questions}
\label{sec:rq}
We answer the following research questions to evaluate the performance of the studied neural code generation models on ML programming tasks.

\begin{itemize}
 \item  \textbf{RQ1 (Accuracy)}: How accurate are the code generation models on different ML programming tasks?

 \item \textbf{RQ2 (Syntactic Correctness)}: How often does the generated code for ML programming tasks pass syntactic checking?

 \item  \textbf{RQ3 (Semantic Correctness)}: How often is the generated code for ML programming tasks pass semantic checking?

 \item  \textbf{RQ4 (Usefulness)}: Is the generated code for ML programming tasks useful for developers? 
\end{itemize}


In \textbf{RQ1}, we set out to investigate the performance of state-of-the-art neural code generation models on the three different types of ML programming tasks. 
In \textbf{RQ2} and \textbf{RQ3}, we assess the soundness of the generated code by checking its syntactic and semantic correctness, respectively. 
In \textbf{RQ4}, we explore the usefulness of the generated code to developers in practice via a user case study.

\subsection{Experiment Setup}
\label{sec:expSetup}

The selected six neural code generation models are required to apply specific data pre-processing steps, i.e., building AST from the source code used in \textbf{tranX}~\cite{yin2018tranx}. Thus to rebuild these models, we first apply the data pre-processing scripts provided in their replication packages. 
For training each model, we select 80\% of our training data to train the model and the left 20\% are used as validation data. 

To maximize the potential of the evaluated approaches, we perform hyper-parameter tuning for each of the evaluated approaches. 
For \textbf{tranX}, we use 0.3 for the dropout, a hidden size of 256 and an embedding size of 128, a learning rate of 0.001, batch size of 64.
We train the model for a maximum of 20 epochs and use the beam size of 15 when testing.
For \textbf{EK-codegen}, we use the same value of hyper-parameter used in \textbf{tranX}. 
For \textbf{CG-RL}, we used the same dropout, network size, and learning rate for the training.
However, we use 10 epochs to pre-train the branch selector following the original paper of \textbf{CG-RL}~\cite{jiang2021exploring}.
For \textbf{Code-gen-TAE}, we use 1e-05 for encoder learning rate and 7.5e-05 for decoder learning rate, the hidden size of the encoder is 768 with 12 layers, the decoder size is 256 with 4 layers, and the batch size used is 16.
We also trained them for 20 epochs like the rest of the model and used a beam number of 10 when testing.
For \textbf{TreeCodeGen}, 0.1 is used for the dropout, and the hidden and embedding size used is the same as \textbf{tranX}. The learning rate used is 1e-4. The max epoch used for training is 20 and the beam size used for testing is 30.
For \textbf{PyCodeGPT}, the model is trained with 32K vocabularies, 768 for hidden size, 12 for the number of heads and layers, a window size of 256, and an initializer range of 0.02, which is the same settings reported from GPT-Neo and PyCodeGPT. The model was trained for 100K steps as suggested in the original paper.

All the aforementioned settings used for the training are from the papers of the six models and any details missing from the paper are set as the default settings provided by their replication packages. We used the GPUs in the Google Colab Pro \footnote{\url{https://colab.research.google.com/}}, which dedicates either NVIDIA Tesla P100 or T4, for training models.
For training PyCodeGPT, we used NVIDIA A100 40GB from google computing platform\footnote{\url{https://cloud.google.com/}}. {The time cost of model training of these models ranges from 3 hours (i.e., \textbf{tranX}) to 16 hours (i.e., \textbf{PyCodeGPT}).}

\subsection{Evaluation Metrics}
\label{sec:metrics}
To assess the performance of the models in generating high-quality source code, we use the following three evaluation measures, i.e., BLEU~\cite{papineni2002bleu}, METEOR~\cite{banerjee2005meteor}, and ROUGE~\cite{lin2004rouge}.

BLEU score~\cite{papineni2002bleu} is an evaluation metric commonly used in assessing the generation models in natural language processing.
It measures the similarity of the model-generated text to the ground truth text by computing the overlapping n-grams of the two texts.
A value of 0 indicates that the generated source code has no overlap of n-grams with the ground truth source code, while the value of 1 indicates that it is a perfect overlap with the ground truth source code. 
There are different ways to calculate the BLEU score.
We can set a different number of n-grams we want to assess in overlapping, i.e. unigram, bigram, trigram, and 4-gram.
This is also referred as to BLEU-1, BLEU-2, BLEU-3, and BLEU-4 respectively. 
METEOR metric~\cite{banerjee2005meteor} is also commonly used to assess machine language generation models.
The metric is based on the harmonic mean of unigram precision and recall but with a higher weight on recall. 
ROUGE~\cite{lin2004rouge} is a metric used for evaluating natural language generation models to compare the generated sequence to the reference. 
There are also multiple sets of assessing different ROUGE values.
In this study, we report the ROUGE-L value of each model which uses the Longest Common Sub-sequence (LCS) in assessing the similarity of the two texts.
\section{Results and Analysis}
\label{sec:result}
This section presents the experimental results and answers the research questions discussed in Section~\ref{sec:rq}.
\begin{table*}[t!]
\centering \caption{Performance of the studied six neural code generation models on different ML tasks and non-ML tasks. {The best performance of a model on different tasks is shown in bold, while the overall best model on each ML task is shown with asterisk (*).}
	}
	\label{tab:rq1_results}
\begin{tabular}{l|l|l|l|l|l|l|l}
\hline
\textbf{Models} & \textbf{Tasks} & \textbf{BLEU-1} & \textbf{BLEU-2} & \textbf{BLEU-3} & \textbf{BLEU-4} & \textbf{ROUGE-L} & \textbf{METEOR} \\ \hline
\multirow{2}{*}{\textbf{tranX}} &
  \begin{tabular}[c]{@{}l@{}}\emph{data process}\\\emph{model building}\\ \emph{evaluation}\end{tabular} &
  \begin{tabular}[c]{@{}l@{}}0.3606\\ \textbf{0.3646}\\ 0.3621\end{tabular} &
  \begin{tabular}[c]{@{}l@{}}0.2670*\\ \textbf{0.2717}\\ 0.2677\end{tabular} &
  \begin{tabular}[c]{@{}l@{}}0.2237*\\ \textbf{0.2277}\\ 0.2233\end{tabular} &
  \begin{tabular}[c]{@{}l@{}}0.1905\\ \textbf{0.1940}\\ 0.1886\end{tabular} &
  \begin{tabular}[c]{@{}l@{}}0.3380\\ 0.3167\\ \textbf{0.3532}\end{tabular} & 
  \begin{tabular}[c]{@{}l@{}}0.3600*\\ 0.3598\\ \textbf{0.3712}\end{tabular}  \\ \cline{2-8}
  & \emph{non-ml} & 0.2855 & 0.1714 & 0.1236 & 0.0929 & 0.1636 & 0.2282 \\ \hline
\multirow{2}{*}{\textbf{EK Codegen}} &
  \begin{tabular}[c]{@{}l@{}}\emph{data process}\\\emph{model building}\\ \emph{evaluation}\end{tabular} &
  \begin{tabular}[c]{@{}l@{}}0.3391\\ 0.3549\\ \textbf{0.3593} \end{tabular} &
  \begin{tabular}[c]{@{}l@{}}0.2541\\ 0.2690\\ \textbf{0.2698*} \end{tabular} &
  \begin{tabular}[c]{@{}l@{}}0.2154\\ \textbf{0.2289}\\ 0.2282 \end{tabular} &
  \begin{tabular}[c]{@{}l@{}}0.1853\\ \textbf{0.1978}\\ 0.1953 \end{tabular} &
  \begin{tabular}[c]{@{}l@{}}0.3441\\ 0.3241\\ \textbf{0.3616}\end{tabular} &
  \begin{tabular}[c]{@{}l@{}}0.3592\\ 0.3633\\ \textbf{0.3759*}\\ \end{tabular}  \\ \cline{2-8}
  & \emph{non-ml} & 0.2969* & 0.1838* & 0.1348* & 0.1028* & 0.1705 & 0.2365* \\ \hline
\multirow{2}{*}{\textbf{CG-RL}} &
  \begin{tabular}[c]{@{}l@{}}\emph{data process}\\\emph{model building}\\ \emph{evaluation}\end{tabular} &
  \begin{tabular}[c]{@{}l@{}}\textbf{0.2503}\\ 0.1923\\ 0.2404 \end{tabular} &
  \begin{tabular}[c]{@{}l@{}}\textbf{0.1943}\\ 0.1471\\ 0.1840 \end{tabular} &
  \begin{tabular}[c]{@{}l@{}}\textbf{0.1678}\\ 0.1257\\ 0.1575 \end{tabular} &
  \begin{tabular}[c]{@{}l@{}}\textbf{0.1471}\\ 0.1089\\ 0.1366\\ \end{tabular} &
  \begin{tabular}[c]{@{}l@{}}0.3397\\ 0.3160\\ \textbf{0.3620}\\ \end{tabular} &
  \begin{tabular}[c]{@{}l@{}}0.3333\\ 0.3057\\ \textbf{0.3480}\\ \end{tabular} \\ \cline{2-8}
    & \emph{non-ml} & 0.1213 & 0.0776 & 0.0578 & 0.0450 & 0.1723 & 0.1916 \\ \hline
\multirow{2}{*}{\textbf{TreeCodeGen}} &
  \begin{tabular}[c]{@{}l@{}}\emph{data process}\\\emph{model building}\\ \emph{evaluation}\end{tabular} &
  \begin{tabular}[c]{@{}l@{}}0.3465\\ 0.3483\\ \textbf{0.3644*} \end{tabular} &
  \begin{tabular}[c]{@{}l@{}}0.2540\\ 0.2559\\ \textbf{0.2681} \end{tabular} &
  \begin{tabular}[c]{@{}l@{}}0.2125\\ 0.2134\\ \textbf{0.2235} \end{tabular} &
  \begin{tabular}[c]{@{}l@{}}0.1810\\ 0.1812\\ \textbf{0.1883} \end{tabular} &
  \begin{tabular}[c]{@{}l@{}}\textbf{0.3530*}\\ 0.3122\\ \textbf{0.3530} \end{tabular} &
  \begin{tabular}[c]{@{}l@{}}0.3476\\ 0.3377\\ \textbf{0.3619}\\ \end{tabular}  \\ \cline{2-8}
      & \emph{non-ml} & 0.1997 & 0.1255 & 0.0932 & 0.0723 & 0.1929* & 0.1979 \\ \hline
\multirow{2}{*}{\textbf{Codegen-TAE}} &
  \begin{tabular}[c]{@{}l@{}}\emph{data process}\\\emph{model building}\\ \emph{evaluation}\end{tabular} &
  \begin{tabular}[c]{@{}l@{}}\textbf{0.3775*}\\ 0.3279\\ 0.2919 \end{tabular} &
  \begin{tabular}[c]{@{}l@{}}\textbf{0.2444}\\ 0.1993\\ 0.1743 \end{tabular} &
  \begin{tabular}[c]{@{}l@{}}\textbf{0.1857}\\ 0.1438\\ 0.1260 \end{tabular} &
  \begin{tabular}[c]{@{}l@{}}\textbf{0.1441}\\ 0.1052\\ 0.0914 \end{tabular} &
  \begin{tabular}[c]{@{}l@{}}\textbf{0.3145}\\ 0.2339\\ 0.2357 \end{tabular} &
  \begin{tabular}[c]{@{}l@{}}\textbf{0.3233}\\ 0.2623\\ 0.2358 \end{tabular}  \\ \cline{2-8}
        & \emph{non-ml} & 0.2620 & 0.1451 & 0.0985 & 0.0703 & 0.1217  &  0.1944 \\ \hline
\multirow{2}{*}{\textbf{PyCodeGPT}} &
  \begin{tabular}[c]{@{}l@{}}\emph{data process}\\\emph{model building}\\ \emph{evaluation}\end{tabular} &
  \begin{tabular}[c]{@{}l@{}}0.3347\\ \textbf{0.4036*}\\ 0.3397 \end{tabular} &
  \begin{tabular}[c]{@{}l@{}}0.2562\\ \textbf{0.3286*}\\ 0.2693 \end{tabular} &
  \begin{tabular}[c]{@{}l@{}}0.2219\\ \textbf{0.2925*}\\ 0.2361* \end{tabular} &
  \begin{tabular}[c]{@{}l@{}}0.1958*\\ \textbf{0.2634*}\\ 0.2094* \end{tabular} &
  \begin{tabular}[c]{@{}l@{}}0.3333\\ \textbf{0.3701*}\\ 0.3667* \end{tabular} &
  \begin{tabular}[c]{@{}l@{}}0.3462\\ \textbf{0.3893*}\\ 0.3654 \end{tabular}  \\ \cline{2-8}
  & \emph{non-ml} & 0.0789 & 0.0352 & 0.0217 & 0.0141 & 0.0619 & 0.0938 \\ \hline
\end{tabular}
\end{table*}
\subsection{RQ1: Accuracy}
\label{sec:answer_rq1}
\textbf{Approach:}
To answer this RQ, we follow our experiment setups (described in Section~\ref{sec:expSetup}) to re-train each of the six neural code generation models with data from different ML programming tasks, i.e., \emph{data process}, \emph{model building}, and \emph{evaluation} datasets.  
We also evaluate the performance of the six neural code generation models on the non-ML programming task dataset for comparison.
For creating the non-ML programming task dataset, we randomly select samples from JuIce~\cite{agashe2019juice} with ML programming tasks removed. To make a fair comparison, we set the size of the non-ML task dataset to our biggest ML task dataset, i.e. \emph{data process} dataset. To remove potential bias, we repeat the data creation for non-ML tasks 10 times and re-train the models accordingly.
We use the average performance of each studied neural code generation model for comparison.  
To assess the model's capability of generating source code on the different ML libraries, we label each testing instance with the studied ML library using the ML API name. 


\textbf{Result:} 
The performance of the studied six neural code generation models on different types of ML tasks and non-ML tasks are organized in Table~\ref{tab:rq1_results}.
{Overall, \textbf{PyCodeGPT} performs best, especially on ML-related tasks. For example, \textbf{PyCodeGPT} achieves the largest values regarding all the metrics on \emph{model building}. 
However, on non-ML tasks, \textbf{EK Codegen} performs best with 5/6 of the metrics.
A possible reason for these two models to have better performance is that they exploit external data for pre-training. 
Specifically, \textbf{PyCodeGPT} uses corpus mined from GitHub, and \textbf{EK Codegen} uses corpus mined from Python Standard Library \footnote{\url{https://docs.python.org/3/library/index.html}} and StackOverflow\footnote{\url{https://stackoverflow.com/}}.
Due to the abundant number of ML projects on GitHub, \textbf{PyCodeGPT} benefits in generating ML-related code.
Similarly, non-ML tasks are composed of short standalone code snippets, which are very similar to the code snippets answered in StackOverflow. This could be the reason why \textbf{EK Codegen} has benefited in generating non-ML programming tasks.}
Regarding the performance on a specific type of ML task, we can see that the best performance tasks for each neural code generation model vary. For example, 
\textbf{tranX} performed best on \emph{model building} tasks over other tasks, while \textbf{CG-RL} and \textbf{Codegen-TAE} have better performance for \emph{data process} tasks than the performance on other tasks regarding BLEU scores. 
\textbf{EK Codegen} and \textbf{TreeCodeGen} perform better on \emph{evaluation}, while \textbf{PyCodeGPT} performs better on \emph{model building} related ML tasks than the other tasks. 

In addition, we can also see that all the six neural code generation models perform significantly better on ML tasks compared to non-ML tasks regarding all the different metrics, i.e., BLEU-1/2/3/4, ROUGE-L, and METEOR. 
Possible reasons for this can be as follows.
First, the non-ML tasks are much more diverse than ML tasks as we limited the ML tasks to only four libraries, i.e., TensorFlow, Keras, Pytorch, and Scikit-learn.
The variance of programs from ML tasks is less than the non-ML dataset. 
Second, the natural language descriptions of the non-ML-related tasks were also more varied and context-free since the domain does not have a constraint compared to the descriptions of ML tasks. 
This finding is interesting, as previous studies state that traditional programs are more deterministic and less statistically orientated~\cite{zhang2020machine}. 
However, from the generation point of view, it is the opposite.
Even though the ML libraries are more dynamic and statistically orientated, programs built upon them are often not. 

\mybox{
\textbf{Good:} Neural code generation models generate significantly better performance on ML tasks than non-ML tasks. \textbf{Bad:} Neural code generation models perform not stable on different ML programming tasks. 
}

\subsection{RQ2: Syntactic Checking}
\label{sec:answer_rq2}
\textbf{Approach:}
To answer this RQ, we conduct syntactic checking on the generated programs for each instance in the testing dataset. 
Following existing work~\cite{liu2020deep}, we conduct a static syntactic checking on the generated programs with the state-of-the-practice tool Pylint\footnote{https://pylint.pycqa.org/}, a Python static code analysis tool that looks for programming errors. 
Note that,  \textbf{codegen-TAE} generates all code statements in one line for a given programming task, while Python programs have strict style requirements, e.g., Python uses indentation to indicate a block of code. To avoid any style errors, we first parse the generated source code and output it with proper indentation. During this step, we do not change any content for avoiding potential bias. 
We then use Pylint to check the syntactic correctness of each instance in the testing dataset, we consider an instance syntactically correct if it passes the checking.


\textbf{Result:}
Table~\ref{tab:rq2_results} shows the results of the syntactic checking for programs generated by different neural code generation models in different ML tasks and non-ML tasks.  
We can observe that in general, all the state-of-the-art models have a high accuracy in generating syntactically correct programs. {However, we observed a significant decline in generating syntactically correct code using \textbf{PyCodeGPT} compared to other models.} 
One of the possible reasons for this is that these five models are AST-based approaches, given a programming task, they first generate the corresponding AST and then transfer them into a code snippet.
{On the other hand, \textbf{PyCodeGPT} uses a token-based seq2seq Transformer, giving the model the freedom to generate any token sequences, which can be easily syntactic incorrect.}
In addition, we can also see that all the six neural code generation models perform better on most of the ML tasks compared to non-ML tasks, which is consistent with our findings in RQ1. 
\mybox{\textbf{Good:} The neural code generation models generate highly accurate syntactically correct programs for ML tasks. \textbf{Good:} Our results confirm the advantage of AST-based code generation used in current neural code generation models compared to the token-based model.} 

\begin{table}[t!]
\centering
\caption{Syntactical correctness (in percentage) of the generated code by different neural code generation models on different types of machine learning tasks.
}
\label{tab:rq2_results}
\begin{tabular}{|l|l|l|l||l|}
\hline
\textbf{Methods} & \textbf{data} & \textbf{model} & \textbf{evaluation} & \textbf{non-ml} \\ \hline
\textbf{tranX} & 98.40 & 98.68 & 99.36 & 97.06  \\ \hline
\textbf{EK codegen} & 96.39 & 97.92 & 97.68 & 95.16  \\ \hline
\textbf{CG-RL} & 97.37 & 98.07 & 97.97 & 94.70  \\ \hline
\textbf{TreeCodeGen} & 92.24 & 95.46 & 97.00 & 82.85  \\ \hline
\textbf{Codege-TAE} & 92.35 & 93.99 & 90.70 & 93.22  \\ \hline
\textbf{PyCodeGPT} & 71.36 & 70.53 & 72.88 & 67.16  \\ \hline
\end{tabular}
\end{table}

\subsection{RQ3: Semantic Checking}
\label{sec:answer_rq3}
\begin{table}[t!]
\centering
	\caption{Semantic correctness (in percentage) of the generated code by different neural code generation models on different types of machine learning tasks.
	}
	\label{tab:rq3_results}
	\setlength\tabcolsep{6.0pt}
\begin{tabular}{|l|c|c|c||c|}
\hline
\textbf{Methods} & \textbf{data} & \textbf{model} & \textbf{evaluation} & \textbf{non-ml} \\ \hline
\textbf{tranX} & 6\% & 14\% & 34\% & 0\%\\ \hline
\textbf{EK codegn} & 16\% & 12\% & 30\% & 10\%\\ \hline
\textbf{PyCodeGPT} & 34\% & 14\% & 30\% & 6\%\\ \hline
\end{tabular}
\end{table}

\begin{table}[t!]
\centering
	\caption{Correctness of identified APIs on the generated code by different neural code generation models on different types of machine learning tasks.}
	\label{tab:rq3_apis}
\begin{tabular}{|l|c|c|c|c||c|}
\hline
\textbf{Methods} & \textbf{Metrics} & \textbf{data} & \textbf{model} & \textbf{evaluation} & \textbf{non-ml} \\ \hline
\multirow{2}{*}{\textbf{tranX}} &
Precision & 0.2129 & 0.2740 & 0.4348 & 0.1214\\ \cline{2-6} &
Recall & 0.2215 & 0.2729 & 0.5105 & 0.1364\\ \hline
\multirow{2}{*}{\textbf{EK codegen}} &
Precision & 0.3458 & 0.2104 & 0.4234  & 0.2278\\ \cline{2-6} &
Recall & 0.3400 & 0.2693 & 0.4452 & 0.2078\\ \hline
\multirow{2}{*}{\textbf{PyCodeGPT}} &
Precision & 0.6133 & 0.4085 & 0.5500 & 0.0867\\ \cline{2-6} &
Recall & 0.5811 & 0.3943 & 0.4649 & 0.0783\\ \hline
\end{tabular}
\end{table}

\textbf{Approach:} 
To further analyze the soundness of the generated programs by these state-of-the-art neural code generation models, we also check the semantics of the generated programs manually as most of the generated programs require specific inputs to be executed, which cannot be generated automatically.  
For this experiment, we examine the semantic correctness of the generated programs from the best three models, i.e.,  \textbf{tranX}, \textbf{EK codegen}, and \textbf{PyCodeGPT} on the TensorFlow library. 
Specifically, we first randomly select 50 examples from each type of ML programming task, thus in total, we have 450 programming tasks to be checked, i.e., 50 samples $\times$ 3 types of ML programming tasks $\times$ 3 models. 
We consider a generated program of a given programming task is semantically equivalent to the ground-truth program if the generated program shares the same logic with the ground-truth program and can solve the given programming task.
For checking the semantic correctness of each sample, all the authors work together and make the decision together. 

To further illustrate the effectiveness of the generated programs for ML programming tasks, we also evaluate the correctness of a generated program regarding the APIs called in the program. 
Specifically, given an ML programming task, we first parse its corresponding ground truth source code and extract the APIs from these four ML libraries that are used to solve the task.
Then we follow the same process to extract APIs in the generated program for this task by a neural code generation model.  
Note that, for the non-ML tasks, we also follow the same process to collect APIs used in the ground-truth and generated programs.
For measuring the performance of neural code models regarding identifying the correct APIs, we use \texttt{Precision} and \texttt{Recall}, where \texttt{Precision} measures the percentage of correct identified APIs among all the used APIs in the generated programs and \texttt{Recall} is the percentage of correct identified APIs among the ground-truth APIs.


\definecolor{dkgreen}{rgb}{0,0.6,0}
\definecolor{gray}{rgb}{0.5,0.5,0.5}
\definecolor{mauve}{rgb}{0.58,0,0.82}

\lstset{frame=tb,
  language=Python,
  aboveskip=3mm,
  belowskip=3mm,
  showstringspaces=false,
  columns=flexible,
  basicstyle={\small\ttfamily},
  numbers=none,
  numberstyle=\tiny\color{gray},
  keywordstyle=\color{blue},
  commentstyle=\color{dkgreen},
  stringstyle=\color{mauve},
  breaklines=true,
  breakatwhitespace=true,
  tabsize=3
}

\begin{lstlisting}[caption={An example of generated code by \textbf{EK codegen} for a given ML programming task},label={lst:rq3_ex}]
# NL description
# using a feed_dict, same graph,but without 
# hardcoding inputs when building session

# ground truth
a = tf.placeholder(dtype=tf.int32, shape=(None,))
b = tf.placeholder(dtype=tf.int32, shape=(None,))
c = tf.add(a, b)
with tf.Session() as sess:
    result = sess.run(c, feed_dict={
        a: [3, 4, 5], b: [-1, 2, 3]})
    print(result)

# generated
x = tf.placeholder(tf.float32)
with tf.Session() as sess:
    print(sess.run(x))
\end{lstlisting}

\textbf{Result:} Table~\ref{tab:rq3_results} shows the semantic correctness of the generated programs of the three neural code generation models.  
We can observe that only a small number of samples of the generated source code are semantically correct with an average of 17\% out of the 450 samples. {\textbf{PyCodeGPT} outperforms \textbf{EK Codegen} followed by \textbf{tranX} with 13\%, 17\%, and 21\% of average percentage, respectively}, which is consistent with the results from RQ1. 
Similar to the trends revealed in RQ1 and RQ2, the examined models also have better performance on ML programming tasks than non-ML programming tasks. 

Despite the low values in semantic correctness, we observe that these neural code generation models have better performance in suggesting the correct APIs for the given programming tasks compared to directly generating source code. 
Table~\ref{tab:rq3_apis} shows the performance of these models in identifying the correct APIs.
{As we can see, the precision and recall values for identifying correct APIs are higher across all different ML programming tasks, which suggests they can help identify an average of 39\% of correct APIs for solving ML tasks.}
We further show the box plots of precision and recall values regarding identifying correct APIs of the three models in Figure~\ref{fig:precisions} and Figure~\ref{fig:recalls} respectively.

\begin{figure}[t!]
\centering
\begin{subfigure}{0.3\linewidth}
\centering
\scalebox{0.5}{

\begin{tikzpicture}
\pgfplotsset{
	   boxplot/every whisker/.style={thick,solid,black},
	   boxplot/every median/.style={very thick,solid,black},
}
\begin{axis}[
xtick={1,2,3,4,5,6,7,8},
xticklabels= ,
boxplot/draw direction=y,
ymin= -0.1,
ymax= 1.1,
]

\addplot[
very thick,
mark=+,
fill=gray,
boxplot
]
table [row sep=\\, y index=0] {
	data\\  0.0\\ 0.0\\ 0.0\\ 0.0\\ 0.0\\ 0\\ 0.0\\ 0.0\\ 0.0\\ 0\\ 0.0\\ 0.0\\ 0.0\\ 0.0\\ 0.0\\ 0.0\\ 0.0\\ 0.0\\ 0.0\\ 0.0\\ 0.0\\ 0.0\\ 0\\ 0.0\\ 0.0\\ 0\\ 0.0\\ 0.0\\ 0.0\\ 0\\ 0.0\\ 0.1111111111111111\\ 0.2\\ 0.3333333333333333\\ 0.3333333333333333\\ 0.3333333333333333\\ 0.3333333333333333\\ 0.5\\ 0.5\\ 0.5\\ 0.5\\ 0.5\\ 0.5\\ 0.5\\ 0.5\\ 1.0\\ 1.0\\ 1.0\\ 1.0\\ 1.0\\
};

\addplot[
very thick,
mark=+,
fill=white,
boxplot
]
table [row sep=\\, y index=0] {
	data\\0.0\\ 0.0\\ 0.0\\ 0.0\\ 0.0\\ 0.0\\ 0.0\\ 0.0\\ 0.0\\ 0.0\\ 0.0\\ 0.0\\ 0.0\\ 0.0\\ 0.0\\ 0.0\\ 0.0\\ 0.0\\ 0.0\\ 0.0\\ 0.0\\ 0.0\\ 0.0\\ 0.0\\ 0.0\\ 0.0\\ 0.0\\ 0.0\\ 0.0\\ 0.0\\ 0.0\\ 0.0\\ 0.0\\ 0.0\\ 0.0\\ 0.0\\ 0.0\\ 0.0\\ 0.0\\ 0.0\\ 0.16666666666666666\\ 0.16666666666666666\\ 0.2\\ 0.2\\ 0.25\\ 0.25\\ 0.25\\ 0.25\\ 0.2857142857142857\\ 0.2857142857142857\\ 0.2857142857142857\\ 0.3333333333333333\\ 0.3333333333333333\\ 0.3333333333333333\\ 0.3333333333333333\\ 0.3333333333333333\\ 0.4\\ 0.4\\ 0.5\\ 0.5\\ 0.5\\ 0.6666666666666666\\ 0.8\\ 0.8\\ 1.0\\ 1.0\\ 1.0\\ 1.0\\ 1.0\\ 1.0\\ 1.0\\ 1.0\\ 1.0\\ 1.0\\ 1.0\\ 1.0\\
};

\addplot[
very thick,
mark=+,
fill=black,
boxplot
] 
table [row sep=\\, y index=0] {
    data\\ 0.0\\ 0\\ 0.0\\ 0.0\\ 0.0\\ 0.0\\ 0.0\\ 0.0\\ 0.0\\ 0.0\\ 0\\ 0.0\\ 0.0\\ 0.0\\ 0.0\\ 0.0\\ 0.0\\ 0.0\\ 0\\ 0.0\\ 0.0\\ 0.0\\ 0\\ 0.0\\ 0.0\\ 0.0\\ 0.0\\ 0.0\\ 0.09090909090909091\\ 0.125\\ 0.16666666666666666\\ 0.2\\ 0.25\\ 0.3333333333333333\\ 0.3333333333333333\\ 0.3333333333333333\\ 0.3333333333333333\\ 0.4\\ 0.4\\ 0.5\\ 0.5\\ 0.5\\ 0.6666666666666666\\ 1.0\\ 1.0\\ 1.0\\ 1.0\\ 1.0\\ 1.0\\ 1.0\\ 1.0\\ 1.0\\ 1.0\\ 1.0\\ 1.0\\ 1.0\\ 1.0\\ 1.0\\ 1.0\\ 1.0\\ 1.0\\ 1.0\\ 1.0\\ 1.0\\ 1.0\\ 1.0\\ 1.0\\
};

\addplot[
very thick,
mark=+,
boxplot
] 
table [row sep=\\, y index=0] {
    data\\0.0\\ 0\\ 0.0\\ 0.0\\ 0.0\\ 0.0\\ 0.0\\ 0.0\\ 0.0\\ 0.0\\ 0.0\\ 0.0\\ 0.0\\ 0.0\\ 0.0\\ 0.0\\ 0\\ 0\\ 0.0\\ 0.0\\ 0.0\\ 0.1\\ 0.125\\ 0.16666666666666666\\ 0.25\\ 0.3333333333333333\\ 0.5\\ 0.5\\ 0.6666666666666666\\ 1.0\\
};

\end{axis}
\end{tikzpicture}
}
\vspace{-0.1in}
\caption{\small tranX}
\end{subfigure}
\begin{subfigure}{0.3\linewidth}
\centering
\scalebox{0.5}{

\begin{tikzpicture}
\pgfplotsset{
	   boxplot/every whisker/.style={thick,solid,black},
	   boxplot/every median/.style={very thick,solid,black},
}
\begin{axis}[
xtick={1,2,3,4,5,6,7,8},
xticklabels= ,
boxplot/draw direction=y,
ymin= -0.1,
ymax= 1.1,
]

\addplot[
very thick,
mark=+,
fill=gray,
boxplot
]
table [row sep=\\, y index=0] {
	data\\ 0.0\\ 0.0\\ 0.0\\ 0.0\\ 0\\ 0.0\\ 0.0\\ 0.0\\ 0.0\\ 0.0\\ 0.0\\ 0.0\\ 0.0\\ 0.0\\ 0.0\\ 0.0\\ 0.0\\ 0.0\\ 0.0\\ 0.0\\ 0.0\\ 0.0\\ 0\\ 0.0\\ 0.0\\ 0.0\\ 0.125\\ 0.25\\ 0.25\\ 0.3333333333333333\\ 0.3333333333333333\\ 0.5\\ 0.5\\ 0.5\\ 0.5\\ 0.5\\ 0.5\\ 1.0\\ 1.0\\ 1.0\\ 1.0\\ 1.0\\ 1.0\\ 1.0\\ 1.0\\ 1.0\\ 1.0\\ 1.0\\ 1.0\\ 1.0\\
};

\addplot[
very thick,
mark=+,
fill=white,
boxplot
]
table [row sep=\\, y index=0] {
	data\\0.0\\ 0.0\\ 0.0\\ 0.0\\ 0.0\\ 0.0\\ 0.0\\ 0.0\\ 0.0\\ 0.0\\ 0.0\\ 0.0\\ 0.0\\ 0.0\\ 0.0\\ 0.0\\ 0.0\\ 0.0\\ 0.0\\ 0.0\\ 0.0\\ 0.0\\ 0.0\\ 0.0\\ 0.0\\ 0.0\\ 0.0\\ 0.0\\ 0.0\\ 0.0\\ 0.0\\ 0.0\\ 0.0\\ 0.0\\ 0.0\\ 0.0\\ 0.0\\ 0.0\\ 0.0\\ 0.0\\ 0.0\\ 0.0\\ 0.1\\ 0.14285714285714285\\ 0.14285714285714285\\ 0.16666666666666666\\ 0.2\\ 0.2\\ 0.2\\ 0.2\\ 0.2\\ 0.2\\ 0.2\\ 0.2222222222222222\\ 0.2222222222222222\\ 0.25\\ 0.25\\ 0.25\\ 0.3333333333333333\\ 0.3333333333333333\\ 0.3333333333333333\\ 0.3333333333333333\\ 0.4\\ 0.4\\ 0.5\\ 1.0\\ 1.0\\ 1.0\\ 1.0\\ 1.0\\ 1.0\\ 1.0\\ 1.0\\ 1.0\\ 1.0\\
};

\addplot[
very thick,
mark=+,
fill=black,
boxplot
] 
table [row sep=\\, y index=0] {
    data\\0.0\\ 0.0\\ 0.0\\ 0.0\\ 0.0\\ 0.0\\ 0.0\\ 0.0\\ 0.0\\ 0.0\\ 0.0\\ 0.0\\ 0.0\\ 0.0\\ 0.0\\ 0.0\\ 0.0\\ 0\\ 0.0\\ 0.0\\ 0\\ 0.0\\ 0.0\\ 0.0\\ 0.0\\ 0.0\\ 0.0\\ 0.0\\ 0.0\\ 0.16666666666666666\\ 0.25\\ 0.3333333333333333\\ 0.3333333333333333\\ 0.3333333333333333\\ 0.3333333333333333\\ 0.3333333333333333\\ 0.5\\ 0.5\\ 0.5\\ 0.6666666666666666\\ 1.0\\ 1.0\\ 1.0\\ 1.0\\ 1.0\\ 1.0\\ 1.0\\ 1.0\\ 1.0\\ 1.0\\ 1.0\\ 1.0\\ 1.0\\ 1.0\\ 1.0\\ 1.0\\ 1.0\\ 1.0\\ 1.0\\ 1.0\\ 1.0\\ 1.0\\
};
\addplot[
very thick,
mark=+,
boxplot
] 
table [row sep=\\, y index=0] {
    data\\0.0\\ 0.0\\ 0.0\\ 0.0\\ 0.0\\ 0.0\\ 0\\ 0.0\\ 0.0\\ 0.0\\ 0.0\\ 0.0\\ 0.0\\ 0.0\\ 0\\ 0.0\\ 0\\ 0.0\\ 0\\ 0.0\\ 0\\ 0\\ 0.3333333333333333\\ 0.5\\ 1.0\\ 1.0\\ 1.0\\ 1.0\\ 1.0\\ 1.0\\
};
\end{axis}
\end{tikzpicture}

}
\vspace{-0.1in}
\caption{\small EK Codegen}
\end{subfigure}
\begin{subfigure}{0.3\linewidth}
\centering
\scalebox{0.5}{

\begin{tikzpicture}
\pgfplotsset{
	   boxplot/every whisker/.style={thick,solid,black},
	   boxplot/every median/.style={very thick,solid,black},
}
\begin{axis}[
xtick={1,2,3,4,5,6,7,8},
xticklabels= ,
boxplot/draw direction=y,
ymin= -0.1,
ymax= 1.1,
]

\addplot[
very thick,
mark=+,
fill=gray,
boxplot
]
table [row sep=\\, y index=0] {
	data\\0.0\\ 0.0\\ 0.0\\ 0.0\\ 0.0\\ 0.0\\ 0.0\\ 0.0\\ 0.0\\ 0.0\\ 0.0\\ 0.0\\ 0.0\\ 0.16666666666666666\\ 0.5\\ 0.5\\ 0.5\\ 0.5\\ 0.5\\ 0.5\\ 0.5\\ 0.5\\ 0.5\\ 0.5\\ 0.5\\ 1.0\\ 1.0\\ 1.0\\ 1.0\\ 1.0\\ 1.0\\ 1.0\\ 1.0\\ 1.0\\ 1.0\\ 1.0\\ 1.0\\ 1.0\\ 1.0\\ 1.0\\ 1.0\\ 1.0\\ 1.0\\ 1.0\\ 1.0\\ 1.0\\ 1.0\\ 1.0\\ 1.0\\ 1.0\\
};

\addplot[
very thick,
mark=+,
fill=white,
boxplot
]
table [row sep=\\, y index=0] {
	data\\ 0.0\\ 0.0\\ 0.0\\ 0.0\\ 0.0\\ 0.0\\ 0.0\\ 0.0\\ 0.0\\ 0.0\\ 0.0\\ 0.0\\ 0.0\\ 0.0\\ 0.0\\ 0.0\\ 0.0\\ 0.0\\ 0.0\\ 0.0\\ 0.14285714285714285\\ 0.2\\ 0.2\\ 0.2\\ 0.25\\ 0.3333333333333333\\ 0.3333333333333333\\ 0.3333333333333333\\ 0.3333333333333333\\ 0.5\\ 0.5\\ 0.5\\ 0.5\\ 0.6666666666666666\\ 0.8\\ 0.8\\ 0.8333333333333334\\ 1.0\\ 1.0\\ 1.0\\ 1.0\\ 1.0\\ 1.0\\ 1.0\\ 1.0\\ 1.0\\ 1.0\\ 1.0\\ 1.0\\ 1.0\\ 
};

\addplot[
very thick,
mark=+,
fill=black,
boxplot
] 
table [row sep=\\, y index=0] {
data\\0.0\\ 0.0\\ 0.0\\ 0.0\\ 0.0\\ 0.0\\ 0.0\\ 0.0\\ 0.0\\ 0.0\\ 0.0\\ 0.0\\ 0.0\\ 0\\ 0.0\\ 0.0\\ 0.0\\ 0.0\\ 0.0\\ 0.0\\ 0.0\\ 0.5\\ 0.5\\ 0.5\\ 1.0\\ 1.0\\ 1.0\\ 1.0\\ 1.0\\ 1.0\\ 1.0\\ 1.0\\ 1.0\\ 1.0\\ 1.0\\ 1.0\\ 1.0\\ 1.0\\ 1.0\\ 1.0\\ 1.0\\ 1.0\\ 1.0\\ 1.0\\ 1.0\\ 1.0\\ 1.0\\ 1.0\\ 1.0\\ 1.0\\ 
};

\addplot[
very thick,
mark=+,
boxplot
] 
table [row sep=\\, y index=0] {
    data\\0\\ 0\\ 0\\ 0.0\\ 0.0\\ 0.0\\ 0.0\\ 0.0\\ 0.0\\ 0.0\\ 0.0\\ 0\\ 0.0\\ 0\\ 0.0\\ 0.0\\ 0\\ 0.0\\ 0.0\\ 0\\ 0\\ 0\\ 0\\ 0\\ 0\\ 0.0\\ 0\\ 0\\ 0.0\\ 0.0\\ 0\\ 0\\ 0\\ 0.0\\ 0\\ 0.0\\ 0\\ 0\\ 0.0\\ 0\\ 0\\ 0\\ 0\\ 0.0\\ 0.0\\ 0.3333333333333333\\ 1.0\\ 1.0\\ 1.0\\ 1.0\\
};

\end{axis}
\end{tikzpicture}

}
\vspace{-0.1in}
\caption{\small PyCodeGPT}
\end{subfigure}

\caption{Precision of identifying correct APIs of \textbf{tranX} and \textbf{EK Codegen} \textbf{PyCodeGPT}, on \emph{data}, \emph{model}, \emph{eval}, \emph{nonml}, respectively.}
\label{fig:precisions}
\end{figure}
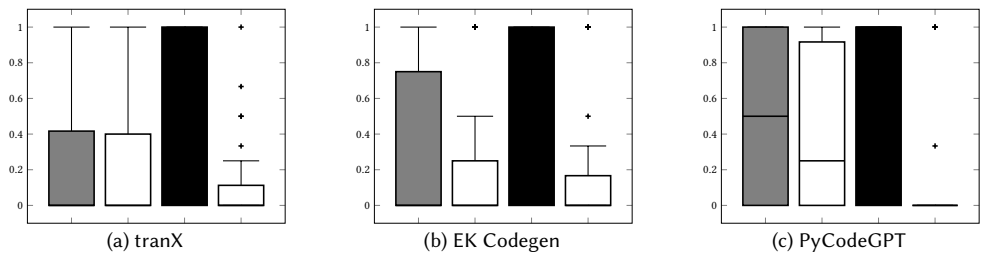
\begin{figure}[t!]
\centering
\begin{subfigure}{0.32\linewidth}
\centering
\scalebox{0.5}{

\begin{tikzpicture}
\pgfplotsset{
	   boxplot/every whisker/.style={thick,solid,black},
	   boxplot/every median/.style={very thick,solid,black},
}
\begin{axis}[
xtick={1,2,3,4,5,6,7,8},
xticklabels= ,
boxplot/draw direction=y,
ymin= -0.1,
ymax= 1.1,
]

\addplot[
very thick,
mark=+,
fill=gray,
boxplot
]
table [row sep=\\, y index=0] {
	data\\ 0.0\\ 0.0\\ 0.0\\ 0.0\\ 0.0\\ 0\\ 0.0\\ 0.0\\ 0.0\\ 0\\ 0.0\\ 0.0\\ 0.0\\ 0.0\\ 0.0\\ 0.0\\ 0.0\\ 0.0\\ 0.0\\ 0.0\\ 0.0\\ 0.0\\ 0\\ 0.0\\ 0.0\\ 0\\ 0.0\\ 0.0\\ 0.0\\ 0\\ 0.0\\ 0.125\\ 0.2\\ 0.25\\ 0.3333333333333333\\ 0.3333333333333333\\ 0.3333333333333333\\ 0.5\\ 0.5\\ 0.5\\ 0.5\\ 0.5\\ 0.5\\ 0.5\\ 1.0\\ 1.0\\ 1.0\\ 1.0\\ 1.0\\ 1.0\\
};

\addplot[
very thick,
mark=+,
fill=white,
boxplot
]
table [row sep=\\, y index=0] {
	data\\0.0\\ 0.0\\ 0.0\\ 0.0\\ 0.0\\ 0.0\\ 0.0\\ 0.0\\ 0.0\\ 0.0\\ 0.0\\ 0.0\\ 0.0\\ 0.0\\ 0.0\\ 0.0\\ 0.0\\ 0.0\\ 0.0\\ 0.0\\ 0.0\\ 0.0\\ 0.0\\ 0.0\\ 0.0\\ 0.0\\ 0.0\\ 0.0\\ 0.0\\ 0.0\\ 0.0\\ 0.0\\ 0.0\\ 0.0\\ 0.0\\ 0.0\\ 0.0\\ 0.0\\ 0.0\\ 0.0\\ 0.125\\ 0.16666666666666666\\ 0.16666666666666666\\ 0.16666666666666666\\ 0.16666666666666666\\ 0.2222222222222222\\ 0.2222222222222222\\ 0.2222222222222222\\ 0.25\\ 0.25\\ 0.25\\ 0.25\\ 0.25\\ 0.3333333333333333\\ 0.3333333333333333\\ 0.3333333333333333\\ 0.3333333333333333\\ 0.3333333333333333\\ 0.3333333333333333\\ 0.42857142857142855\\ 0.8\\ 0.8\\ 1.0\\ 1.0\\ 1.0\\ 1.0\\ 1.0\\ 1.0\\ 1.0\\ 1.0\\ 1.0\\ 1.0\\ 1.0\\ 1.0\\ 1.0\\ 1.0\\ 
};

\addplot[
very thick,
mark=+,
fill=black,
boxplot
] 
table [row sep=\\, y index=0] {
    data\\0.0\\ 0\\ 0.0\\ 0.0\\ 0.0\\ 0.0\\ 0.0\\ 0.0\\ 0.0\\ 0.0\\ 0\\ 0.0\\ 0.0\\ 0.0\\ 0.0\\ 0.0\\ 0.0\\ 0.0\\ 0\\ 0.0\\ 0.0\\ 0.0\\ 0\\ 0.0\\ 0.0\\ 0.0\\ 0.0\\ 0.0\\ 0.25\\ 0.2857142857142857\\ 0.3333333333333333\\ 0.3333333333333333\\ 0.5\\ 0.5\\ 0.5\\ 0.5\\ 1.0\\ 1.0\\ 1.0\\ 1.0\\ 1.0\\ 1.0\\ 1.0\\ 1.0\\ 1.0\\ 1.0\\ 1.0\\ 1.0\\ 1.0\\ 1.0\\ 1.0\\ 1.0\\ 1.0\\ 1.0\\ 1.0\\ 1.0\\ 1.0\\ 1.0\\ 1.0\\ 1.0\\ 1.0\\ 1.0\\ 1.0\\ 1.0\\ 1.0\\ 1.0\\ 1.0\\ 
};

\addplot[
very thick,
mark=+,
boxplot
] 
table [row sep=\\, y index=0] {
    data\\ 0.0\\ 0\\ 0.0\\ 0.0\\ 0.0\\ 0.0\\ 0.0\\ 0.0\\ 0.0\\ 0.0\\ 0.0\\ 0.0\\ 0.0\\ 0.0\\ 0.0\\ 0.0\\ 0\\ 0\\ 0.0\\ 0.0\\ 0.0\\ 0.14285714285714285\\ 0.2\\ 0.25\\ 0.3333333333333333\\ 0.5\\ 0.5\\ 0.5\\ 0.6666666666666666\\ 1.0\\
};

\end{axis}
\end{tikzpicture}

}
\vspace{-0.1in}
\caption{\small tranX}
\end{subfigure}
\begin{subfigure}{0.32\linewidth}
\centering
\scalebox{0.5}{

\begin{tikzpicture}
\pgfplotsset{
	   boxplot/every whisker/.style={thick,solid,black},
	   boxplot/every median/.style={very thick,solid,black},
}
\begin{axis}[
xtick={1,2,3,4,5,6,7,8},
xticklabels= ,
boxplot/draw direction=y,
ymin= -0.1,
ymax= 1.1,
]

\addplot[
very thick,
mark=+,
fill=gray,
boxplot
]
table [row sep=\\, y index=0] {
	data\\0.0\\ 0.0\\ 0.0\\ 0.0\\ 0\\ 0.0\\ 0.0\\ 0.0\\ 0.0\\ 0.0\\ 0.0\\ 0.0\\ 0.0\\ 0.0\\ 0.0\\ 0.0\\ 0.0\\ 0.0\\ 0.0\\ 0.0\\ 0.0\\ 0.0\\ 0\\ 0.0\\ 0.0\\ 0.0\\ 0.25\\ 0.25\\ 0.3333333333333333\\ 0.3333333333333333\\ 0.3333333333333333\\ 0.3333333333333333\\ 0.5\\ 0.5\\ 0.5\\ 0.5\\ 0.5\\ 0.6666666666666666\\ 1.0\\ 1.0\\ 1.0\\ 1.0\\ 1.0\\ 1.0\\ 1.0\\ 1.0\\ 1.0\\ 1.0\\ 1.0\\ 1.0\\
};

\addplot[
very thick,
mark=+,
fill=white,
boxplot
]
table [row sep=\\, y index=0] {
	data\\ 0.0\\ 0.0\\ 0.0\\ 0.0\\ 0.0\\ 0.0\\ 0.0\\ 0.0\\ 0.0\\ 0.0\\ 0.0\\ 0.0\\ 0.0\\ 0.0\\ 0.0\\ 0.0\\ 0.0\\ 0.0\\ 0.0\\ 0.0\\ 0.0\\ 0.0\\ 0.0\\ 0.0\\ 0.0\\ 0.0\\ 0.0\\ 0.0\\ 0.0\\ 0.0\\ 0.0\\ 0.0\\ 0.0\\ 0.0\\ 0.0\\ 0.0\\ 0.0\\ 0.0\\ 0.0\\ 0.0\\ 0.0\\ 0.0\\ 0.09090909090909091\\ 0.14285714285714285\\ 0.16666666666666666\\ 0.16666666666666666\\ 0.16666666666666666\\ 0.2\\ 0.25\\ 0.25\\ 0.3333333333333333\\ 0.3333333333333333\\ 0.3333333333333333\\ 0.3333333333333333\\ 0.3333333333333333\\ 0.3333333333333333\\ 0.3333333333333333\\ 0.5\\ 0.6\\ 0.6666666666666666\\ 0.6666666666666666\\ 1.0\\ 1.0\\ 1.0\\ 1.0\\ 1.0\\ 1.0\\ 1.0\\ 1.0\\ 1.0\\ 1.0\\ 1.0\\ 1.0\\ 1.0\\ 1.0\\
};

\addplot[
very thick,
mark=+,
fill=black,
boxplot
]
table [row sep=\\, y index=0] {
    data\\ 0\\ 0.0\\ 0.0\\ 0.0\\ 0.0\\ 0.0\\ 0.0\\ 0.0\\ 0.0\\ 0.0\\ 0.0\\ 0.0\\ 0.0\\ 0.0\\ 0.0\\ 0.0\\ 0.0\\ 0\\ 0.0\\ 0.0\\ 0\\ 0.0\\ 0.0\\ 0.0\\ 0.0\\ 0.0\\ 0.0\\ 0.0\\ 0.0\\ 0.06666666666666667\\ 0.25\\ 0.2857142857142857\\ 0.3333333333333333\\ 0.3333333333333333\\ 0.3333333333333333\\ 0.5\\ 0.5\\ 1.0\\ 1.0\\ 1.0\\ 1.0\\ 1.0\\ 1.0\\ 1.0\\ 1.0\\ 1.0\\ 1.0\\ 1.0\\ 1.0\\ 1.0\\ 1.0\\ 1.0\\ 1.0\\ 1.0\\ 1.0\\ 1.0\\ 1.0\\ 1.0\\ 1.0\\ 1.0\\ 1.0\\ 1.0\\
};

\addplot[
very thick,
mark=+,
boxplot
] 
table [row sep=\\, y index=0] {
    data\\0.0\\ 0.0\\ 0.0\\ 0.0\\ 0.0\\ 0.0\\ 0\\ 0.0\\ 0.0\\ 0.0\\ 0.0\\ 0.0\\ 0.0\\ 0.0\\ 0\\ 0.0\\ 0\\ 0.0\\ 0\\ 0.0\\ 0\\ 0\\ 0.3333333333333333\\ 0.4\\ 0.5\\ 1.0\\ 1.0\\ 1.0\\ 1.0\\ 1.0\\
};

\end{axis}
\end{tikzpicture}

}
\vspace{-0.1in}
\caption{\small EK codegen}
\end{subfigure}
\begin{subfigure}{0.32\linewidth}
\centering
\scalebox{0.5}{

\begin{tikzpicture}
\pgfplotsset{
	   boxplot/every whisker/.style={thick,solid,black},
	   boxplot/every median/.style={very thick,solid,black},
}
\begin{axis}[
xtick={1,2,3,4,5,6,7,8},
xticklabels= ,
boxplot/draw direction=y,
ymin= -0.1,
ymax= 1.1,
]

\addplot[
very thick,
mark=+,
fill=gray,
boxplot
]
table [row sep=\\, y index=0] {
	data\\0.0\\ 0.0\\ 0.0\\ 0.0\\ 0.0\\ 0.0\\ 0.0\\ 0.0\\ 0.0\\ 0.0\\ 0.0\\ 0.0\\ 0.0\\ 0.2\\ 0.3333333333333333\\ 0.3333333333333333\\ 0.3333333333333333\\ 0.3333333333333333\\ 0.3333333333333333\\ 0.3333333333333333\\ 0.5\\ 0.5\\ 0.5\\ 0.5\\ 0.5\\ 0.5\\ 0.8571428571428571\\ 1.0\\ 1.0\\ 1.0\\ 1.0\\ 1.0\\ 1.0\\ 1.0\\ 1.0\\ 1.0\\ 1.0\\ 1.0\\ 1.0\\ 1.0\\ 1.0\\ 1.0\\ 1.0\\ 1.0\\ 1.0\\ 1.0\\ 1.0\\ 1.0\\ 1.0\\ 1.0\\
};

\addplot[
very thick,
mark=+,
fill=white,
boxplot
]
table [row sep=\\, y index=0] {
	data\\ 0.0\\ 0.0\\ 0.0\\ 0.0\\ 0.0\\ 0.0\\ 0.0\\ 0.0\\ 0.0\\ 0.0\\ 0.0\\ 0.0\\ 0.0\\ 0.0\\ 0.0\\ 0.0\\ 0.0\\ 0.0\\ 0.0\\ 0.0\\ 0.1111111111111111\\ 0.125\\ 0.14285714285714285\\ 0.16666666666666666\\ 0.2\\ 0.25\\ 0.3333333333333333\\ 0.45454545454545453\\ 0.5\\ 0.5\\ 0.5\\ 0.5\\ 0.5\\ 0.6666666666666666\\ 0.6666666666666666\\ 0.6666666666666666\\ 0.8\\ 0.8\\ 0.8333333333333334\\ 1.0\\ 1.0\\ 1.0\\ 1.0\\ 1.0\\ 1.0\\ 1.0\\ 1.0\\ 1.0\\ 1.0\\ 1.0\\
};

\addplot[
very thick,
mark=+,
fill=black,
boxplot
]
table [row sep=\\, y index=0] {
    data\\0.0\\ 0.0\\ 0.0\\ 0.0\\ 0.0\\ 0.0\\ 0.0\\ 0.0\\ 0.0\\ 0.0\\ 0.0\\ 0.0\\ 0.0\\ 0\\ 0.0\\ 0.0\\ 0.0\\ 0.0\\ 0.0\\ 0.0\\ 0.0\\ 0.14285714285714285\\ 0.16666666666666666\\ 0.3333333333333333\\ 0.3333333333333333\\ 0.3333333333333333\\ 0.5\\ 0.5\\ 0.6\\ 0.75\\ 0.8333333333333334\\ 0.875\\ 0.875\\ 1.0\\ 1.0\\ 1.0\\ 1.0\\ 1.0\\ 1.0\\ 1.0\\ 1.0\\ 1.0\\ 1.0\\ 1.0\\ 1.0\\ 1.0\\ 1.0\\ 1.0\\ 1.0\\ 1.0\\
};

\addplot[
very thick,
mark=+,
boxplot
] 
table [row sep=\\, y index=0] {
    data\\0\\ 0\\ 0\\ 0.0\\ 0.0\\ 0.0\\ 0.0\\ 0.0\\ 0.0\\ 0.0\\ 0.0\\ 0\\ 0.0\\ 0\\ 0.0\\ 0.0\\ 0\\ 0.0\\ 0.0\\ 0\\ 0\\ 0\\ 0\\ 0\\ 0\\ 0.0\\ 0\\ 0\\ 0.0\\ 0.0\\ 0\\ 0\\ 0\\ 0.0\\ 0\\ 0.0\\ 0\\ 0\\ 0.0\\ 0\\ 0\\ 0\\ 0\\ 0.0\\ 0.0\\ 0.25\\ 0.6666666666666666\\ 1.0\\ 1.0\\ 1.0\\
};

\end{axis}
\end{tikzpicture}

}
\vspace{-0.1in}
\caption{\small PyCodeGPT}
\end{subfigure}

\caption{Recall of identifying correct APIs of \textbf{tranX} and \textbf{EK codegen} \textbf{PyCodeGPT}, on \emph{data}, \emph{model}, \emph{eval}, \emph{nonml}, respectively.}
\label{fig:recalls}
\end{figure}
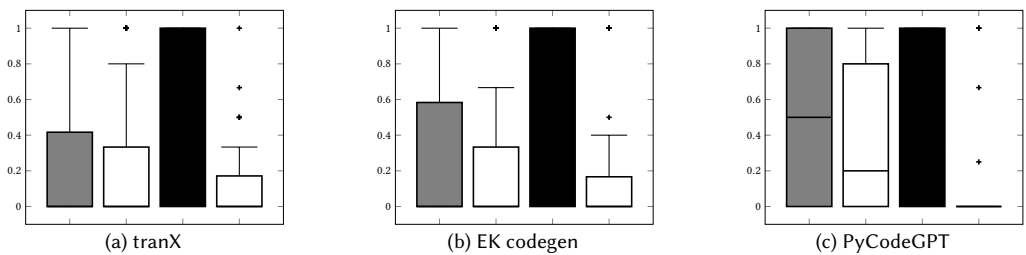


We further show an example ML task that cannot be solved by existing neural code generation models while some of the required APIs can be identified in Listing~\ref{lst:rq3_ex}.
The task is labeled as a \emph{model building} task finished with the TensorFlow library, 
which is about building and running a session by loading a dictionary object.
The ground truth code instantiates two placeholders with a \textit{tf.int32} type and adds them before running a session.
When they run the session, they feed a dictionary object before running them.
While the generated source code failed to generate the dictionary part, it successfully generates code that instantiates a \textit{tf.int32} typed placeholder and then builds and runs a session successfully. 
The precision and recall regarding identifying the correct APIs are 100\% and 57\% respectively.


\mybox{\textbf{Bad:} All the examined neural code generation models are weak in generating semantically correct programs. \textbf{Good:} The generated programs have the potential to be used for identifying correct APIs for ML programming tasks.
}  

\subsection{RQ4: Usefulness}
\label{sec:answer_rq4}

\textbf{Approach:}
To explore the practical value of neural code generation models, we further conduct a user case study to investigate
whether programs generated by these neural code models can help developers finish given programming tasks efficiently and accurately. In this experiment, we examine the programs generated by the best two models, i.e., \textbf{PyCodeGPT} and \textbf{EK codegen} on the TensorFlow library. We randomly selected 20 programming tasks from the test dataset of TensorFlow. We invited 12 students (i.e., six Ph.D. students and six MS students) who are familiar with the TensorFlow framework to complete the 20 programming tasks. Their experience in developing ML tasks based on TensorFlow varied from two to four years, with an average of three years.
We then divided the participants into two groups (Group1 and Group2) based on their experience. 
The 20 tasks were also randomly divided into two groups (Task1 and Task2). The experiment was conducted in two steps. In the first step, Group1 and Group2 were asked to work on Task1 while Group1 will be given the generated program from both \textbf{PyCodeGPT} and \textbf{EK codegen}, in the second step, Group1 and Group2 were asked to work on Task2 while Group2 will be given the generated program from both \textbf{PyCodeGPT} and \textbf{EK codegen}. Each participant was asked to record
his/her screen during the experiment for getting the time he/she spent on solving each programming task. 
Following existing studies~\cite{liu2020deep,xie2020api}, we use correctness and completion time to measure the performance of the participants in solving ML tasks. 
Specifically, correctness evaluates whether a participant can write the correct code for a given task. Note that, as developers might solve the same task with different logic, thus we measure the proportion of correct APIs submitted by a participant among all APIs in the ground truth answer of the task. 
Completion time evaluates how quickly a participant can solve a given task. 
For each task, we recorded the correctness and completion time of each participant.

\begin{figure}[t!]
\centering
\begin{subfigure}{0.44\linewidth}
\centering
\scalebox{0.5}{
\begin{tikzpicture}
\pgfplotsset{
	   boxplot/every whisker/.style={thick,solid,black},
	   boxplot/every median/.style={very thick,solid,black},
}
\begin{axis}[
xtick={1,2,3},
xticklabels={ 
\texttt{Without},,\texttt{With}
},
boxplot/draw direction=y,
ymin= 0,
ymax= 1,
]

\addplot[
very thick,
mark=+,
fill=gray,
boxplot
]
table [row sep=\\, y index=0] {
	data\\0.4\\0.5\\0.65\\0.6\\0.8\\0.34\\0.25\\0.53\\0.44\\0.5\\
};

\addplot[
very thick,
mark=+,
fill=white,
boxplot
]
table [row sep=\\, y index=0] {
	data\\0\\
};

\addplot[
very thick,
mark=+,
boxplot] 
table [row sep=\\, y index=0] {
data\\0.73\\.65\\0.4\\0.6\\0.6\\0.9\\0.45\\0.6\\0.7\\0.55\\
};

\end{axis}
\end{tikzpicture}
}
\vspace{-0.1in}
\caption{\small Correctness}
\end{subfigure}
\begin{subfigure}{0.44\linewidth}
\centering
\scalebox{0.5}{
\begin{tikzpicture}
\pgfplotsset{
	   boxplot/every whisker/.style={thick,solid,black},
	   boxplot/every median/.style={very thick,solid,black},
}
\begin{axis}[
xtick={1,2,3},
xticklabels={ 
\texttt{Without},,\texttt{With}
},
boxplot/draw direction=y,
ymin= 2,
ymax= 10,
]

\addplot[
very thick,
mark=+,
fill=gray,
boxplot
]
table [row sep=\\, y index=0] {
	data\\4\\3\\6.8\\8\\5\\2.4\\7.8\\6.9\\9.0\\7.4\\
};

\addplot[
very thick,
mark=,
fill=white,
boxplot
]
table [row sep=\\, y index=0] {
	data\\0\\
};

\addplot[
very thick,
mark=+,
boxplot] 
table [row sep=\\, y index=0] {
data\\5.4\\5.25\\6.4\\5.1\\6.6\\7.3\\7\\8\\3.2\\2.3\\
};
\end{axis}
\end{tikzpicture}
}
\vspace{-0.1in}
\caption{\small Completion Time (mins)}
\end{subfigure}

\caption{Performance of participants with (denoted as `\textbf{With}' in the figure) and without (denoted as `\textbf{Without}' in the figure) the help of code generation models.}
\label{fig:usercase}
\end{figure}
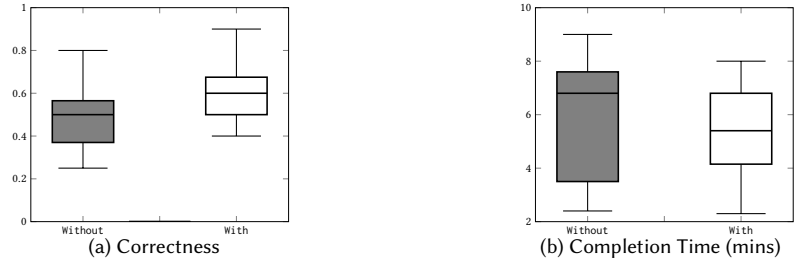

\textbf{Result:}
Figure~\ref{fig:usercase} shows the distribution of correctness and completion time of participants to finish the given 20 ML tasks. Overall, with the generated code from the two code generation models, participants completed the programming tasks more accurately and took slightly less time than participants without any clues. On average, the correctness and completion time of participants with and without the generated code are 0.62 and 5.2, versus 0.51 and 5.6 respectively. We further used Wilcoxon signed-rank test for verifying the statistical significance of the differences. The p-value for correctness is small than 0.05 while the p-value for completion time is larger than 0.05. The result suggests that the generated code cannot significantly reduce the time cost of ML programming tasks, while it can help developers write more correct code. 
One possible reason for this is that most of the generated programs are semantic incorrect, modifying the generated programs is not easy and requires a great amount of time.
Our discussion with the participants confirmed that the examined code generation models can help find a part of the correct APIs for ML tasks, which provides developers with useful clues, thus further improving developers' correctness.


\mybox{\textbf{Bad:} Code generation models cannot significantly improve developers' completion time as most of the generated programs are semantic incorrect, which requires a non-trivial time for modifications. \textbf{Good:} The generated code can help developers write more correct code by providing developers with clues for using correct APIs. 
}
\section{Discussion}
\label{sec:discussion}


\subsection{The Missing of Code Generation on ML Tasks}
\label{sec:dis_issue2}

The above results and analysis have revealed several good and bad aspects of state-of-the-art neural code generation models on ML programming tasks. This section further summarizes the missing aspect that can serve as the practical guidelines for improving code generation tasks for ML tasks.

\vspace{4pt}
\noindent \textbf{Decompose code generation for divide-and-conquer.}
Even powered with state-of-the-art deep neural techniques, code generation for ML programming tasks is still a challenge.
Directly generating code from a given NL-described programming task often produces incorrect and incompetent programs that cannot be used by developers. 
Nevertheless, the findings of our user case study reveal that incorrectly generated code could still be useful for developers as far as it contains essential APIs to solve the tasks.
Meanwhile, identifying the correct APIs is easier than generating the correct programs directly.
Taken in this sense, future code generation techniques for ML tasks can be decomposed into two sequential tasks, i.e., API sequence identification and API usage generation.
Crowd-sourced knowledge from Stack Overflow can be used to help generate the context of the identified API sequences, which has been proved to be useful in mining API usage scenarios~\cite{uddin2020mining, huang2018api}.

\vspace{4pt}
\noindent \textbf{Human-machine collaborated code generation.}
The examined state-of-the-art neural code generation models treat code generation as a machine translation task and generate statements/blocks from a given NL-described programming task sequentially, during which any incorrect statement could mislead the generation of the sequential code snippets. 
Meanwhile, during our case study,  similar to the findings of Murali et al.~\cite{murali2018neural}, we also find that, despite lacking the whole knowledge for certain programming tasks, developers usually know whether it is correct for some generated statements, or whether a specific API should be utilized based on their knowledge of an ML library.  
If human knowledge could be integrated into the process of automatic code generation, e.g., in the form of simple feedback from end-users to assess the correctness of the intermediate uncertain statements, the code generation could be significantly improved.

Such practice has another advantage, i.e. personalize the generated code that fit the knowledge of different end-users, by integrating developers' knowledge into the code generation process. 

\vspace{4pt}
\noindent \textbf{Focus on domain-specific tasks.}
Compared to the general non-ML programming tasks, all of the six neural code generation models exert better performance regarding syntactic correctness, semantic correctness, and identifying essential APIs, on domain-specific tasks, i.e., ML programming tasks.
We assume this might be because the domain-specific task only involves a smaller set of APIs (indicating little knowledge needed to be learned), thus training an effective model is relatively easier compared to the model training on a dataset with diversified tasks.
We also notice that previous practices on code generation were usually conducted on general tasks~\cite{yao2018staqc,yin2018learning,oda2015learning,agashe2019juice}. 
Thus, future code generation should focus on training or fine-tuning models with data from specific domains to improve the performance of generating programs for domain-specific tasks, which have much more potential than general programming tasks.
We also encourage the need for the design of domain-specific models by incorporating the characteristics of the domain, which might further improve the performance of specific tasks.


\subsection{Threats to Validity}
\label{sec:threats}

\noindent \textbf{Internal Validity.} The main threat to the internal validity of our study is the limited number of models that are compared.
Due to this limitation, we can not generalize our findings from this paper to all the source code generation models.
However, we have used the most recently published and the state-of-the-art performing ones to conduct our experiments, we also described a detailed methodology and the setup of the study and the data used, which will allow future researchers to replicate our study and further explore other source code generation models. 

\noindent \textbf{External Validity.} The main threat to external validity is the labeling of ground-truth data explored in the study.
Although we have put much effort to maintain high-quality data by using real-world programs contributed by developers on GitHub, it could not be enough to validate the implementation of the source code.
Also, programming or implementing a programming task can lead to many different forms and logic of source code. Thus, these programs in our dataset might not represent all the ways to implement a given programming task. 

\noindent \textbf{Construct Validity.} The main threat to the construct validity can be the evaluation metrics we used.
BLEU, ROUGE-L, and METEOR scores can assess program generation by calculating how the generated programs of neural code generation models are similar to the ground-truth programs of given tasks in our dataset. 
However as mentioned from the external validity, other implementations can be a reasonable answer rather than using similarity to the ground truth.
In our future study, we plan to examine a different set of evaluation metrics to assess the generation of the program. 

\section{Related Work}
\label{sec:related}
\subsection{Automatic Code Generation}
\label{sec:atuo_codegeneration}
Besides the six neural code generation models used in this paper, there are many other methods for automatic code generation from natural language descriptions proposed in the recent decade~\cite{yin2017syntactic,ling2016latent,rabinovich2017abstract}.
More specifically, deep neural networks in the natural language processing field have been increasingly adopted to encapsulate the complexity of generating general-purpose programming languages such as Java and Python.
Dong and Lapata \cite{dong2018coarse} proposed \emph{Coarse2Fine} which uses a structure-aware neural architecture that applies two decoding steps of semantic parsing. First, the approach decodes a rough sketch of the meaning representation with a high level where it omits the details such as identifiers.
Then they apply another mechanism that fills the missing details by conditioning the natural language description together with the sketch of the general meaning representation.
Hayati et al. \cite{hayati2018retrieval} proposed \emph{ReCode} which exploits sub-tree retrieval for explicitly referencing code examples within a neural code generation model. This approach adopts a retrieval-based neural machine translation approach in the context of source code generation. This is the first approach that introduced the retrieval technique to code generation. 
Murali et al. \cite{murali2018neural} proposed an approach that combines neural network and combinatorial search to generate source code from sketches, which is an abstract style of source code that masks user-defined constants and identifiers. 
Sun et al. \cite{sun2019grammar} proposed a grammar-based structural convolutional neural network (CNN) that utilizes tree-based convolution, pre-order convolution, and attentive pooling layers for a more compact prediction than seq2seq models. The tree-based convolution applies a sliding window on the AST structures. Pre-order convolution that traverses the nodes of the partial AST. These two modules of CNN help capture the neighbor information of the trees. The attentive pooling is designed to aggregate the CNN features to predict better identifier names.
Sun et al. \cite{sun2020treegen} proposed \emph{TreeGen} which is a tree-based neural architecture that uses the attention mechanism of Transformers to alleviate the long dependency issues and utilizes an AST encoder to incorporate grammar rules and AST structures into the network. They utilized a convolution sub-layer only on the first Transformer decoder blocks to prevent the Transformer to mix information from other nodes which may leak the original information.
Lyu et al. \cite{lyu2021embedding} proposed a generation model that exploits the dependencies of the API method as an API dependency graph and utilizes the graph embedding into a seq2seq model. They incorporate an encoder-decoder module that captures both the global structural dependencies and textual program descriptions when generating the source code.
Ahmad et al. \cite{ahmad2021unified} proposed \emph{PLBART}, which is an auto-regressive transformer model that pre-trains on an extensive collection of unlabeled Java and Python functions that are paired with natural language texts through denoising auto-encoders. This approach helps reduce the effort of mining highly human-curated annotations when fine-tuning specific tasks. They have evaluated \emph{PLBART} on several tasks, i.e. code summarization, code generation, code translation, program repair, clone detection, and vulnerability detection.

API recommendation is {can be seen as a kind of} code generation, which targets {to generate} a specific API set for a given query from developers~\cite{mcmillan2011portfolio, chan2012searching, raghothaman2016swim, rahman2016rack, liu2018effective, xie2020api, chen2021holistic, chen2022more}.
Liu et al. \cite{liu2018effective} proposed a method called \emph{RecRank} that applies a novel ranking-based discriminative approach that leverages API usage path features to improve their performance on the recommendation.
Xie et al. \cite{xie2020api} studied the relationship between verb phrase patterns that describes the functionality of APIs and the user queries for finding those APIs. 
Chen et al. \cite{chen2021holistic} proposed a novel API recommendation method that combines API usage with the text information in the source code based on an API context graph network.
Chen et al. \cite{chen2022more} 
propose COOK, which is an API sequence recommendation that combines deep learning models and post-processing strategies.
They have enhanced the beam search with code-specific heuristics that helped them improve the quality of the recommendation.


\subsection{Analysis of Model Effectiveness}
\label{modelEff}
There also have been numerous empirical studies that investigated the effectiveness of source code generation models in different contexts.
Liu et al. \cite{liu2020deep} investigated the effectiveness of source code generation models when natural language requirement texts of general programming are given. They have built a large-scale dataset, \emph{ReCa}, to assess five source code generation models in handling longer sequences of requirement texts.
Chirkova and Troshin \cite{chirkova2021empirical} conducted an empirical study on approaches that apply Transformers \cite{vaswani2017attention} to source code in the software engineering domain. They investigated how Transformers utilize the syntactic information of source code in modeling code completion, function naming, and bug fixing tasks.
Antonio et al. \cite{mastropaolo2021studying} empirically assessed the effect of T5 (Text-to-Text Transfer Transformer) \cite{raffel2020exploring} architecture when applied to source code. They found that their T5 model outperforms four existing baselines when bigger data is used in pre-training.
Peng et al. \cite{peng2022revisiting} did a study to unify the tasks and benchmark datasets of API recommendations to fully facilitate future studies. They grouped the different approaches concerning the input they handle (i.e. query or code). They propose a general benchmark dataset, \emph{APIBENCH}, to assess 11 existing approaches in both query and code-based API recommendations.

These empirical studies assess the effectiveness of source code generation models in generating general programming tasks with respect to different levels of NL complexity \cite{liu2020deep} and different architecture \cite{chirkova2021empirical, mastropaolo2021studying}.
The difference in our work is that we investigate the effectiveness of generation models in generating machine/deep learning-related source code using ML libraries.

\section{Conclusion}
\label{sec:conclusion}
This paper conducts an empirical study to explore the performance of six state-of-the-art neural code generation models on ML programming tasks.
A new benchmark dataset that contains 83K pairs of natural language described programming tasks and the corresponding programs implemented by four ML libraries is introduced. 
Our empirical study reveals some good, bad, and missing aspects, with a few major ones listed below. 
(\textbf{Good}) Neural code generation models perform significantly better on ML tasks than on non-ML tasks.
(\textbf{Bad}) Most of the generated code is semantically incorrect. 
{(\textbf{Bad}) Code generation models cannot significantly improve developers’ completion time. (\textbf{Good}) The generated code can help developers write more correct code by providing developers with clues for using correct APIs.} 
(\textbf{Missing}) The observation from our user study reveals the missing aspects of code generation for ML tasks, e.g., decomposing code generation for divide-and-conquer into two tasks: API sequence identification and API usage generation.


\bibliographystyle{ACM-Reference-Format}
\bibliography{reference}


\begin{thebibliography}{58}


\ifx \showCODEN    \undefined \def \showCODEN     #1{\unskip}     \fi
\ifx \showDOI      \undefined \def \showDOI       #1{#1}\fi
\ifx \showISBNx    \undefined \def \showISBNx     #1{\unskip}     \fi
\ifx \showISBNxiii \undefined \def \showISBNxiii  #1{\unskip}     \fi
\ifx \showISSN     \undefined \def \showISSN      #1{\unskip}     \fi
\ifx \showLCCN     \undefined \def \showLCCN      #1{\unskip}     \fi
\ifx \shownote     \undefined \def \shownote      #1{#1}          \fi
\ifx \showarticletitle \undefined \def \showarticletitle #1{#1}   \fi
\ifx \showURL      \undefined \def \showURL       {\relax}        \fi
\providecommand\bibfield[2]{#2}
\providecommand\bibinfo[2]{#2}
\providecommand\natexlab[1]{#1}
\providecommand\showeprint[2][]{arXiv:#2}

\bibitem[Agashe et~al\mbox{.}(2019)]%
        {agashe2019juice}
\bibfield{author}{\bibinfo{person}{Rajas Agashe}, \bibinfo{person}{Srinivasan
  Iyer}, {and} \bibinfo{person}{Luke Zettlemoyer}.}
  \bibinfo{year}{2019}\natexlab{}.
\newblock \showarticletitle{Juice: A large scale distantly supervised dataset
  for open domain context-based code generation}.
\newblock \bibinfo{journal}{\emph{arXiv preprint arXiv:1910.02216}}
  (\bibinfo{year}{2019}).
\newblock


\bibitem[Ahmad et~al\mbox{.}(2021a)]%
        {ahmad2021unified}
\bibfield{author}{\bibinfo{person}{Wasi Ahmad}, \bibinfo{person}{Saikat
  Chakraborty}, \bibinfo{person}{Baishakhi Ray}, {and} \bibinfo{person}{Kai-Wei
  Chang}.} \bibinfo{year}{2021}\natexlab{a}.
\newblock \showarticletitle{Unified Pre-training for Program Understanding and
  Generation}. In \bibinfo{booktitle}{\emph{Proceedings of the 2021 Conference
  of the North American Chapter of the Association for Computational
  Linguistics: Human Language Technologies}}. \bibinfo{pages}{2655--2668}.
\newblock


\bibitem[Ahmad et~al\mbox{.}(2021b)]%
        {ahmad2021avatar}
\bibfield{author}{\bibinfo{person}{Wasi~Uddin Ahmad},
  \bibinfo{person}{Md~Golam~Rahman Tushar}, \bibinfo{person}{Saikat
  Chakraborty}, {and} \bibinfo{person}{Kai-Wei Chang}.}
  \bibinfo{year}{2021}\natexlab{b}.
\newblock \showarticletitle{AVATAR: A Parallel Corpus for Java-Python Program
  Translation}.
\newblock \bibinfo{journal}{\emph{arXiv preprint arXiv:2108.11590}}
  (\bibinfo{year}{2021}).
\newblock


\bibitem[Amershi et~al\mbox{.}(2019)]%
        {amershi2019software}
\bibfield{author}{\bibinfo{person}{Saleema Amershi}, \bibinfo{person}{Andrew
  Begel}, \bibinfo{person}{Christian Bird}, \bibinfo{person}{Robert DeLine},
  \bibinfo{person}{Harald Gall}, \bibinfo{person}{Ece Kamar},
  \bibinfo{person}{Nachiappan Nagappan}, \bibinfo{person}{Besmira Nushi}, {and}
  \bibinfo{person}{Thomas Zimmermann}.} \bibinfo{year}{2019}\natexlab{}.
\newblock \showarticletitle{Software engineering for machine learning: A case
  study}. In \bibinfo{booktitle}{\emph{2019 IEEE/ACM 41st International
  Conference on Software Engineering: Software Engineering in Practice
  (ICSE-SEIP)}}. IEEE, \bibinfo{pages}{291--300}.
\newblock


\bibitem[Banerjee and Lavie(2005)]%
        {banerjee2005meteor}
\bibfield{author}{\bibinfo{person}{Satanjeev Banerjee} {and}
  \bibinfo{person}{Alon Lavie}.} \bibinfo{year}{2005}\natexlab{}.
\newblock \showarticletitle{METEOR: An automatic metric for MT evaluation with
  improved correlation with human judgments}. In
  \bibinfo{booktitle}{\emph{Proceedings of the acl workshop on intrinsic and
  extrinsic evaluation measures for machine translation and/or summarization}}.
  \bibinfo{pages}{65--72}.
\newblock


\bibitem[Black et~al\mbox{.}(2021)]%
        {black2021gpt}
\bibfield{author}{\bibinfo{person}{Sid Black}, \bibinfo{person}{Leo Gao},
  \bibinfo{person}{Phil Wang}, \bibinfo{person}{Connor Leahy}, {and}
  \bibinfo{person}{Stella Biderman}.} \bibinfo{year}{2021}\natexlab{}.
\newblock \showarticletitle{GPT-Neo: Large Scale Autoregressive Language
  Modeling with Mesh-Tensorflow}.
\newblock \bibinfo{journal}{\emph{If you use this software, please cite it
  using these metadata}}  \bibinfo{volume}{58} (\bibinfo{year}{2021}).
\newblock


\bibitem[Brown et~al\mbox{.}(2020)]%
        {brown2020language}
\bibfield{author}{\bibinfo{person}{Tom Brown}, \bibinfo{person}{Benjamin Mann},
  \bibinfo{person}{Nick Ryder}, \bibinfo{person}{Melanie Subbiah},
  \bibinfo{person}{Jared~D Kaplan}, \bibinfo{person}{Prafulla Dhariwal},
  \bibinfo{person}{Arvind Neelakantan}, \bibinfo{person}{Pranav Shyam},
  \bibinfo{person}{Girish Sastry}, \bibinfo{person}{Amanda Askell},
  {et~al\mbox{.}}} \bibinfo{year}{2020}\natexlab{}.
\newblock \showarticletitle{Language models are few-shot learners}.
\newblock \bibinfo{journal}{\emph{Advances in neural information processing
  systems}}  \bibinfo{volume}{33} (\bibinfo{year}{2020}),
  \bibinfo{pages}{1877--1901}.
\newblock


\bibitem[Bui et~al\mbox{.}(2021)]%
        {bui2021self}
\bibfield{author}{\bibinfo{person}{Nghi~DQ Bui}, \bibinfo{person}{Yijun Yu},
  {and} \bibinfo{person}{Lingxiao Jiang}.} \bibinfo{year}{2021}\natexlab{}.
\newblock \showarticletitle{Self-Supervised Contrastive Learning for Code
  Retrieval and Summarization via Semantic-Preserving Transformations}. In
  \bibinfo{booktitle}{\emph{Proceedings of the 44th International ACM SIGIR
  Conference on Research and Development in Information Retrieval}}.
  \bibinfo{pages}{511--521}.
\newblock


\bibitem[Chan et~al\mbox{.}(2012)]%
        {chan2012searching}
\bibfield{author}{\bibinfo{person}{Wing-Kwan Chan}, \bibinfo{person}{Hong
  Cheng}, {and} \bibinfo{person}{David Lo}.} \bibinfo{year}{2012}\natexlab{}.
\newblock \showarticletitle{Searching connected API subgraph via text phrases}.
  In \bibinfo{booktitle}{\emph{Proceedings of the ACM SIGSOFT 20th
  International Symposium on the Foundations of Software Engineering}}.
  \bibinfo{pages}{1--11}.
\newblock


\bibitem[Chen et~al\mbox{.}(2022)]%
        {chen2022more}
\bibfield{author}{\bibinfo{person}{Chi Chen}, \bibinfo{person}{Xin Peng},
  \bibinfo{person}{Bihuan Chen}, \bibinfo{person}{Jun Sun},
  \bibinfo{person}{Zhenchang Xing}, \bibinfo{person}{Xin Wang}, {and}
  \bibinfo{person}{Wenyun Zhao}.} \bibinfo{year}{2022}\natexlab{}.
\newblock \showarticletitle{“More Than Deep Learning”: post-processing for
  API sequence recommendation}.
\newblock \bibinfo{journal}{\emph{Empirical Software Engineering}}
  \bibinfo{volume}{27}, \bibinfo{number}{1} (\bibinfo{year}{2022}),
  \bibinfo{pages}{1--32}.
\newblock


\bibitem[Chen et~al\mbox{.}(2021)]%
        {chen2021holistic}
\bibfield{author}{\bibinfo{person}{Chi Chen}, \bibinfo{person}{Xin Peng},
  \bibinfo{person}{Zhenchang Xing}, \bibinfo{person}{Jun Sun},
  \bibinfo{person}{Xin Wang}, \bibinfo{person}{Yifan Zhao}, {and}
  \bibinfo{person}{Wenyun Zhao}.} \bibinfo{year}{2021}\natexlab{}.
\newblock \showarticletitle{Holistic combination of structural and textual code
  information for context based API recommendation}.
\newblock \bibinfo{journal}{\emph{IEEE Transactions on Software Engineering}}
  (\bibinfo{year}{2021}).
\newblock


\bibitem[Chirkova and Troshin(2021)]%
        {chirkova2021empirical}
\bibfield{author}{\bibinfo{person}{Nadezhda Chirkova} {and}
  \bibinfo{person}{Sergey Troshin}.} \bibinfo{year}{2021}\natexlab{}.
\newblock \showarticletitle{Empirical study of transformers for source code}.
  In \bibinfo{booktitle}{\emph{Proceedings of the 29th ACM Joint Meeting on
  European Software Engineering Conference and Symposium on the Foundations of
  Software Engineering}}. \bibinfo{pages}{703--715}.
\newblock


\bibitem[Cozzie and King(2012)]%
        {cozzie2012macho}
\bibfield{author}{\bibinfo{person}{Anthony~E Cozzie} {and}
  \bibinfo{person}{Samuel King}.} \bibinfo{year}{2012}\natexlab{}.
\newblock \showarticletitle{Macho: Writing programs with natural language and
  examples}.
\newblock  (\bibinfo{year}{2012}).
\newblock


\bibitem[Dahal et~al\mbox{.}(2021)]%
        {dahal2021analysis}
\bibfield{author}{\bibinfo{person}{Samip Dahal}, \bibinfo{person}{Adyasha
  Maharana}, {and} \bibinfo{person}{Mohit Bansal}.}
  \bibinfo{year}{2021}\natexlab{}.
\newblock \showarticletitle{Analysis of Tree-Structured Architectures for Code
  Generation}. In \bibinfo{booktitle}{\emph{Findings of the Association for
  Computational Linguistics: ACL-IJCNLP 2021}}. \bibinfo{pages}{4382--4391}.
\newblock


\bibitem[Devlin et~al\mbox{.}(2018)]%
        {devlin2018bert}
\bibfield{author}{\bibinfo{person}{Jacob Devlin}, \bibinfo{person}{Ming-Wei
  Chang}, \bibinfo{person}{Kenton Lee}, {and} \bibinfo{person}{Kristina
  Toutanova}.} \bibinfo{year}{2018}\natexlab{}.
\newblock \showarticletitle{Bert: Pre-training of deep bidirectional
  transformers for language understanding}.
\newblock \bibinfo{journal}{\emph{arXiv preprint arXiv:1810.04805}}
  (\bibinfo{year}{2018}).
\newblock


\bibitem[Dong and Lapata(2016)]%
        {dong2016language}
\bibfield{author}{\bibinfo{person}{Li Dong} {and} \bibinfo{person}{Mirella
  Lapata}.} \bibinfo{year}{2016}\natexlab{}.
\newblock \showarticletitle{Language to logical form with neural attention}.
\newblock \bibinfo{journal}{\emph{arXiv preprint arXiv:1601.01280}}
  (\bibinfo{year}{2016}).
\newblock


\bibitem[Dong and Lapata(2018)]%
        {dong2018coarse}
\bibfield{author}{\bibinfo{person}{Li Dong} {and} \bibinfo{person}{Mirella
  Lapata}.} \bibinfo{year}{2018}\natexlab{}.
\newblock \showarticletitle{Coarse-to-Fine Decoding for Neural Semantic
  Parsing}. In \bibinfo{booktitle}{\emph{Proceedings of the 56th Annual Meeting
  of the Association for Computational Linguistics (Volume 1: Long Papers)}}.
  \bibinfo{pages}{731--742}.
\newblock


\bibitem[Gatt and Krahmer(2018)]%
        {gatt2018survey}
\bibfield{author}{\bibinfo{person}{Albert Gatt} {and} \bibinfo{person}{Emiel
  Krahmer}.} \bibinfo{year}{2018}\natexlab{}.
\newblock \showarticletitle{Survey of the state of the art in natural language
  generation: Core tasks, applications and evaluation}.
\newblock \bibinfo{journal}{\emph{Journal of Artificial Intelligence Research}}
   \bibinfo{volume}{61} (\bibinfo{year}{2018}), \bibinfo{pages}{65--170}.
\newblock


\bibitem[Gu et~al\mbox{.}(2018)]%
        {gu2018deep}
\bibfield{author}{\bibinfo{person}{Xiaodong Gu}, \bibinfo{person}{Hongyu
  Zhang}, {and} \bibinfo{person}{Sunghun Kim}.}
  \bibinfo{year}{2018}\natexlab{}.
\newblock \showarticletitle{Deep code search}. In
  \bibinfo{booktitle}{\emph{2018 IEEE/ACM 40th International Conference on
  Software Engineering (ICSE)}}. IEEE, \bibinfo{pages}{933--944}.
\newblock


\bibitem[Guo et~al\mbox{.}(2020)]%
        {guo2020graphcodebert}
\bibfield{author}{\bibinfo{person}{Daya Guo}, \bibinfo{person}{Shuo Ren},
  \bibinfo{person}{Shuai Lu}, \bibinfo{person}{Zhangyin Feng},
  \bibinfo{person}{Duyu Tang}, \bibinfo{person}{Shujie Liu},
  \bibinfo{person}{Long Zhou}, \bibinfo{person}{Nan Duan},
  \bibinfo{person}{Alexey Svyatkovskiy}, \bibinfo{person}{Shengyu Fu},
  {et~al\mbox{.}}} \bibinfo{year}{2020}\natexlab{}.
\newblock \showarticletitle{Graphcodebert: Pre-training code representations
  with data flow}.
\newblock \bibinfo{journal}{\emph{arXiv preprint arXiv:2009.08366}}
  (\bibinfo{year}{2020}).
\newblock


\bibitem[Hayati et~al\mbox{.}(2018)]%
        {hayati2018retrieval}
\bibfield{author}{\bibinfo{person}{Shirley~Anugrah Hayati},
  \bibinfo{person}{Raphael Olivier}, \bibinfo{person}{Pravalika Avvaru},
  \bibinfo{person}{Pengcheng Yin}, \bibinfo{person}{Anthony Tomasic}, {and}
  \bibinfo{person}{Graham Neubig}.} \bibinfo{year}{2018}\natexlab{}.
\newblock \showarticletitle{Retrieval-Based Neural Code Generation}. In
  \bibinfo{booktitle}{\emph{Proceedings of the 2018 Conference on Empirical
  Methods in Natural Language Processing}}.
\newblock


\bibitem[Huang and Ling(2005)]%
        {huang2005using}
\bibfield{author}{\bibinfo{person}{Jin Huang} {and} \bibinfo{person}{Charles~X
  Ling}.} \bibinfo{year}{2005}\natexlab{}.
\newblock \showarticletitle{Using AUC and accuracy in evaluating learning
  algorithms}.
\newblock \bibinfo{journal}{\emph{IEEE Transactions on knowledge and Data
  Engineering}} \bibinfo{volume}{17}, \bibinfo{number}{3}
  (\bibinfo{year}{2005}), \bibinfo{pages}{299--310}.
\newblock


\bibitem[Huang et~al\mbox{.}(2018)]%
        {huang2018api}
\bibfield{author}{\bibinfo{person}{Qiao Huang}, \bibinfo{person}{Xin Xia},
  \bibinfo{person}{Zhenchang Xing}, \bibinfo{person}{David Lo}, {and}
  \bibinfo{person}{Xinyu Wang}.} \bibinfo{year}{2018}\natexlab{}.
\newblock \showarticletitle{API method recommendation without worrying about
  the task-API knowledge gap}. In \bibinfo{booktitle}{\emph{2018 33rd IEEE/ACM
  International Conference on Automated Software Engineering (ASE)}}. IEEE,
  \bibinfo{pages}{293--304}.
\newblock


\bibitem[Jiang et~al\mbox{.}(2021)]%
        {jiang2021exploring}
\bibfield{author}{\bibinfo{person}{Hui Jiang}, \bibinfo{person}{Chulun Zhou},
  \bibinfo{person}{Fandong Meng}, \bibinfo{person}{Biao Zhang},
  \bibinfo{person}{Jie Zhou}, \bibinfo{person}{Degen Huang},
  \bibinfo{person}{Qingqiang Wu}, {and} \bibinfo{person}{Jinsong Su}.}
  \bibinfo{year}{2021}\natexlab{}.
\newblock \showarticletitle{Exploring Dynamic Selection of Branch Expansion
  Orders for Code Generation}. In \bibinfo{booktitle}{\emph{Proceedings of the
  59th Annual Meeting of the Association for Computational Linguistics and the
  11th International Joint Conference on Natural Language Processing (Volume 1:
  Long Papers)}}. \bibinfo{pages}{5076--5085}.
\newblock


\bibitem[Jozefowicz et~al\mbox{.}(2016)]%
        {jozefowicz2016exploring}
\bibfield{author}{\bibinfo{person}{Rafal Jozefowicz}, \bibinfo{person}{Oriol
  Vinyals}, \bibinfo{person}{Mike Schuster}, \bibinfo{person}{Noam Shazeer},
  {and} \bibinfo{person}{Yonghui Wu}.} \bibinfo{year}{2016}\natexlab{}.
\newblock \showarticletitle{Exploring the limits of language modeling}.
\newblock \bibinfo{journal}{\emph{arXiv preprint arXiv:1602.02410}}
  (\bibinfo{year}{2016}).
\newblock


\bibitem[Khalid et~al\mbox{.}(2014)]%
        {khalid2014survey}
\bibfield{author}{\bibinfo{person}{Samina Khalid}, \bibinfo{person}{Tehmina
  Khalil}, {and} \bibinfo{person}{Shamila Nasreen}.}
  \bibinfo{year}{2014}\natexlab{}.
\newblock \showarticletitle{A survey of feature selection and feature
  extraction techniques in machine learning}. In \bibinfo{booktitle}{\emph{2014
  science and information conference}}. IEEE, \bibinfo{pages}{372--378}.
\newblock


\bibitem[LeCun et~al\mbox{.}(2015)]%
        {lecun2015deep}
\bibfield{author}{\bibinfo{person}{Yann LeCun}, \bibinfo{person}{Yoshua
  Bengio}, {and} \bibinfo{person}{Geoffrey Hinton}.}
  \bibinfo{year}{2015}\natexlab{}.
\newblock \showarticletitle{Deep learning}.
\newblock \bibinfo{journal}{\emph{nature}} \bibinfo{volume}{521},
  \bibinfo{number}{7553} (\bibinfo{year}{2015}), \bibinfo{pages}{436--444}.
\newblock


\bibitem[Lin(2004)]%
        {lin2004rouge}
\bibfield{author}{\bibinfo{person}{Chin-Yew Lin}.}
  \bibinfo{year}{2004}\natexlab{}.
\newblock \showarticletitle{Rouge: A package for automatic evaluation of
  summaries}. In \bibinfo{booktitle}{\emph{Text summarization branches out}}.
  \bibinfo{pages}{74--81}.
\newblock


\bibitem[Ling et~al\mbox{.}(2016)]%
        {ling2016latent}
\bibfield{author}{\bibinfo{person}{Wang Ling}, \bibinfo{person}{Phil Blunsom},
  \bibinfo{person}{Edward Grefenstette}, \bibinfo{person}{Karl~Moritz Hermann},
  \bibinfo{person}{Tom{\'a}{\v{s}} Ko{\v{c}}isk{\`y}}, \bibinfo{person}{Fumin
  Wang}, {and} \bibinfo{person}{Andrew Senior}.}
  \bibinfo{year}{2016}\natexlab{}.
\newblock \showarticletitle{Latent Predictor Networks for Code Generation}. In
  \bibinfo{booktitle}{\emph{Proceedings of the 54th Annual Meeting of the
  Association for Computational Linguistics (Volume 1: Long Papers)}}.
  \bibinfo{pages}{599--609}.
\newblock


\bibitem[Liu et~al\mbox{.}(2016)]%
        {liu2016neural}
\bibfield{author}{\bibinfo{person}{Chang Liu}, \bibinfo{person}{Xin Wang},
  \bibinfo{person}{Richard Shin}, \bibinfo{person}{Joseph~E Gonzalez}, {and}
  \bibinfo{person}{Dawn Song}.} \bibinfo{year}{2016}\natexlab{}.
\newblock \showarticletitle{Neural code completion}.
\newblock  (\bibinfo{year}{2016}).
\newblock


\bibitem[Liu et~al\mbox{.}(2020)]%
        {liu2020deep}
\bibfield{author}{\bibinfo{person}{Hui Liu}, \bibinfo{person}{Mingzhu Shen},
  \bibinfo{person}{Jiaqi Zhu}, \bibinfo{person}{Nan Niu}, \bibinfo{person}{Ge
  Li}, {and} \bibinfo{person}{Lu Zhang}.} \bibinfo{year}{2020}\natexlab{}.
\newblock \showarticletitle{Deep learning based program generation from
  requirements text: Are we there yet?}
\newblock \bibinfo{journal}{\emph{IEEE Transactions on Software Engineering}}
  (\bibinfo{year}{2020}).
\newblock


\bibitem[Liu et~al\mbox{.}(2018)]%
        {liu2018effective}
\bibfield{author}{\bibinfo{person}{Xiaoyu Liu}, \bibinfo{person}{LiGuo Huang},
  {and} \bibinfo{person}{Vincent Ng}.} \bibinfo{year}{2018}\natexlab{}.
\newblock \showarticletitle{Effective API recommendation without historical
  software repositories}. In \bibinfo{booktitle}{\emph{Proceedings of the 33rd
  ACM/IEEE International Conference on Automated Software Engineering}}.
  \bibinfo{pages}{282--292}.
\newblock


\bibitem[Lyu et~al\mbox{.}(2021)]%
        {lyu2021embedding}
\bibfield{author}{\bibinfo{person}{Chen Lyu}, \bibinfo{person}{Ruyun Wang},
  \bibinfo{person}{Hongyu Zhang}, \bibinfo{person}{Hanwen Zhang}, {and}
  \bibinfo{person}{Songlin Hu}.} \bibinfo{year}{2021}\natexlab{}.
\newblock \showarticletitle{Embedding API dependency graph for neural code
  generation}.
\newblock \bibinfo{journal}{\emph{Empirical Software Engineering}}
  \bibinfo{volume}{26}, \bibinfo{number}{4} (\bibinfo{year}{2021}),
  \bibinfo{pages}{1--51}.
\newblock


\bibitem[Mastropaolo et~al\mbox{.}(2021)]%
        {mastropaolo2021studying}
\bibfield{author}{\bibinfo{person}{Antonio Mastropaolo},
  \bibinfo{person}{Simone Scalabrino}, \bibinfo{person}{Nathan Cooper},
  \bibinfo{person}{David~Nader Palacio}, \bibinfo{person}{Denys Poshyvanyk},
  \bibinfo{person}{Rocco Oliveto}, {and} \bibinfo{person}{Gabriele Bavota}.}
  \bibinfo{year}{2021}\natexlab{}.
\newblock \showarticletitle{Studying the usage of text-to-text transfer
  transformer to support code-related tasks}. In \bibinfo{booktitle}{\emph{2021
  IEEE/ACM 43rd International Conference on Software Engineering (ICSE)}}.
  IEEE, \bibinfo{pages}{336--347}.
\newblock


\bibitem[McMillan et~al\mbox{.}(2011)]%
        {mcmillan2011portfolio}
\bibfield{author}{\bibinfo{person}{Collin McMillan}, \bibinfo{person}{Mark
  Grechanik}, \bibinfo{person}{Denys Poshyvanyk}, \bibinfo{person}{Qing Xie},
  {and} \bibinfo{person}{Chen Fu}.} \bibinfo{year}{2011}\natexlab{}.
\newblock \showarticletitle{Portfolio: finding relevant functions and their
  usage}. In \bibinfo{booktitle}{\emph{Proceedings of the 33rd International
  Conference on Software Engineering}}. \bibinfo{pages}{111--120}.
\newblock


\bibitem[Murali et~al\mbox{.}(2018)]%
        {murali2018neural}
\bibfield{author}{\bibinfo{person}{Vijayaraghavan Murali},
  \bibinfo{person}{Letao Qi}, \bibinfo{person}{Swarat Chaudhuri}, {and}
  \bibinfo{person}{Chris Jermaine}.} \bibinfo{year}{2018}\natexlab{}.
\newblock \showarticletitle{Neural Sketch Learning for Conditional Program
  Generation}. In \bibinfo{booktitle}{\emph{International Conference on
  Learning Representations}}.
\newblock


\bibitem[Nguyen et~al\mbox{.}(2019)]%
        {nguyen2019machine}
\bibfield{author}{\bibinfo{person}{Giang Nguyen}, \bibinfo{person}{Stefan
  Dlugolinsky}, \bibinfo{person}{Martin Bob{\'a}k}, \bibinfo{person}{Viet
  Tran}, \bibinfo{person}{Alvaro Lopez~Garcia}, \bibinfo{person}{Ignacio
  Heredia}, \bibinfo{person}{Peter Mal{\'\i}k}, {and} \bibinfo{person}{Ladislav
  Hluch{\`y}}.} \bibinfo{year}{2019}\natexlab{}.
\newblock \showarticletitle{Machine learning and deep learning frameworks and
  libraries for large-scale data mining: a survey}.
\newblock \bibinfo{journal}{\emph{Artificial Intelligence Review}}
  \bibinfo{volume}{52}, \bibinfo{number}{1} (\bibinfo{year}{2019}),
  \bibinfo{pages}{77--124}.
\newblock


\bibitem[Norouzi et~al\mbox{.}(2021)]%
        {norouzi2021code}
\bibfield{author}{\bibinfo{person}{Sajad Norouzi}, \bibinfo{person}{Keyi Tang},
  {and} \bibinfo{person}{Yanshuai Cao}.} \bibinfo{year}{2021}\natexlab{}.
\newblock \showarticletitle{Code Generation from Natural Language with Less
  Prior Knowledge and More Monolingual Data}. In
  \bibinfo{booktitle}{\emph{Proceedings of the 59th Annual Meeting of the
  Association for Computational Linguistics and the 11th International Joint
  Conference on Natural Language Processing (Volume 2: Short Papers)}}.
  \bibinfo{pages}{776--785}.
\newblock


\bibitem[Oda et~al\mbox{.}(2015)]%
        {oda2015learning}
\bibfield{author}{\bibinfo{person}{Yusuke Oda}, \bibinfo{person}{Hiroyuki
  Fudaba}, \bibinfo{person}{Graham Neubig}, \bibinfo{person}{Hideaki Hata},
  \bibinfo{person}{Sakriani Sakti}, \bibinfo{person}{Tomoki Toda}, {and}
  \bibinfo{person}{Satoshi Nakamura}.} \bibinfo{year}{2015}\natexlab{}.
\newblock \showarticletitle{Learning to generate pseudo-code from source code
  using statistical machine translation}. In \bibinfo{booktitle}{\emph{2015
  30th IEEE/ACM International Conference on Automated Software Engineering
  (ASE)}}. IEEE, \bibinfo{pages}{574--584}.
\newblock


\bibitem[Papineni et~al\mbox{.}(2002)]%
        {papineni2002bleu}
\bibfield{author}{\bibinfo{person}{Kishore Papineni}, \bibinfo{person}{Salim
  Roukos}, \bibinfo{person}{Todd Ward}, {and} \bibinfo{person}{Wei-Jing Zhu}.}
  \bibinfo{year}{2002}\natexlab{}.
\newblock \showarticletitle{Bleu: a method for automatic evaluation of machine
  translation}. In \bibinfo{booktitle}{\emph{Proceedings of the 40th annual
  meeting of the Association for Computational Linguistics}}.
  \bibinfo{pages}{311--318}.
\newblock


\bibitem[Peng et~al\mbox{.}(2022)]%
        {peng2022revisiting}
\bibfield{author}{\bibinfo{person}{Yun Peng}, \bibinfo{person}{Shuqing Li},
  \bibinfo{person}{Wenwei Gu}, \bibinfo{person}{Yichen Li},
  \bibinfo{person}{Wenxuan Wang}, \bibinfo{person}{Cuiyun Gao}, {and}
  \bibinfo{person}{Michael Lyu}.} \bibinfo{year}{2022}\natexlab{}.
\newblock \showarticletitle{Revisiting, Benchmarking and Exploring API
  Recommendation: How Far Are We?}
\newblock \bibinfo{journal}{\emph{IEEE Transactions on Software Engineering}}
  (\bibinfo{year}{2022}).
\newblock


\bibitem[Rabinovich et~al\mbox{.}(2017)]%
        {rabinovich2017abstract}
\bibfield{author}{\bibinfo{person}{Maxim Rabinovich}, \bibinfo{person}{Mitchell
  Stern}, {and} \bibinfo{person}{Dan Klein}.} \bibinfo{year}{2017}\natexlab{}.
\newblock \showarticletitle{Abstract Syntax Networks for Code Generation and
  Semantic Parsing}. In \bibinfo{booktitle}{\emph{Proceedings of the 55th
  Annual Meeting of the Association for Computational Linguistics (Volume 1:
  Long Papers)}}. \bibinfo{pages}{1139--1149}.
\newblock


\bibitem[Raffel et~al\mbox{.}(2020)]%
        {raffel2020exploring}
\bibfield{author}{\bibinfo{person}{Colin Raffel}, \bibinfo{person}{Noam
  Shazeer}, \bibinfo{person}{Adam Roberts}, \bibinfo{person}{Katherine Lee},
  \bibinfo{person}{Sharan Narang}, \bibinfo{person}{Michael Matena},
  \bibinfo{person}{Yanqi Zhou}, \bibinfo{person}{Wei Li},
  \bibinfo{person}{Peter~J Liu}, {et~al\mbox{.}}}
  \bibinfo{year}{2020}\natexlab{}.
\newblock \showarticletitle{Exploring the limits of transfer learning with a
  unified text-to-text transformer.}
\newblock \bibinfo{journal}{\emph{J. Mach. Learn. Res.}} \bibinfo{volume}{21},
  \bibinfo{number}{140} (\bibinfo{year}{2020}), \bibinfo{pages}{1--67}.
\newblock


\bibitem[Raghothaman et~al\mbox{.}(2016)]%
        {raghothaman2016swim}
\bibfield{author}{\bibinfo{person}{Mukund Raghothaman}, \bibinfo{person}{Yi
  Wei}, {and} \bibinfo{person}{Youssef Hamadi}.}
  \bibinfo{year}{2016}\natexlab{}.
\newblock \showarticletitle{Swim: Synthesizing what i mean-code search and
  idiomatic snippet synthesis}. In \bibinfo{booktitle}{\emph{2016 IEEE/ACM 38th
  International Conference on Software Engineering (ICSE)}}. IEEE,
  \bibinfo{pages}{357--367}.
\newblock


\bibitem[Rahman et~al\mbox{.}(2016)]%
        {rahman2016rack}
\bibfield{author}{\bibinfo{person}{Mohammad~Masudur Rahman},
  \bibinfo{person}{Chanchal~K Roy}, {and} \bibinfo{person}{David Lo}.}
  \bibinfo{year}{2016}\natexlab{}.
\newblock \showarticletitle{Rack: Automatic api recommendation using
  crowdsourced knowledge}. In \bibinfo{booktitle}{\emph{2016 IEEE 23rd
  International Conference on Software Analysis, Evolution, and Reengineering
  (SANER)}}, Vol.~\bibinfo{volume}{1}. IEEE, \bibinfo{pages}{349--359}.
\newblock


\bibitem[Sun et~al\mbox{.}(2019)]%
        {sun2019grammar}
\bibfield{author}{\bibinfo{person}{Zeyu Sun}, \bibinfo{person}{Qihao Zhu},
  \bibinfo{person}{Lili Mou}, \bibinfo{person}{Yingfei Xiong},
  \bibinfo{person}{Ge Li}, {and} \bibinfo{person}{Lu Zhang}.}
  \bibinfo{year}{2019}\natexlab{}.
\newblock \showarticletitle{A grammar-based structural cnn decoder for code
  generation}. In \bibinfo{booktitle}{\emph{Proceedings of the AAAI conference
  on artificial intelligence}}, Vol.~\bibinfo{volume}{33}.
  \bibinfo{pages}{7055--7062}.
\newblock


\bibitem[Sun et~al\mbox{.}(2020)]%
        {sun2020treegen}
\bibfield{author}{\bibinfo{person}{Zeyu Sun}, \bibinfo{person}{Qihao Zhu},
  \bibinfo{person}{Yingfei Xiong}, \bibinfo{person}{Yican Sun},
  \bibinfo{person}{Lili Mou}, {and} \bibinfo{person}{Lu Zhang}.}
  \bibinfo{year}{2020}\natexlab{}.
\newblock \showarticletitle{Treegen: A tree-based transformer architecture for
  code generation}. In \bibinfo{booktitle}{\emph{Proceedings of the AAAI
  Conference on Artificial Intelligence}}, Vol.~\bibinfo{volume}{34}.
  \bibinfo{pages}{8984--8991}.
\newblock


\bibitem[Uddin et~al\mbox{.}(2020)]%
        {uddin2020mining}
\bibfield{author}{\bibinfo{person}{Gias Uddin}, \bibinfo{person}{Foutse Khomh},
  {and} \bibinfo{person}{Chanchal~K Roy}.} \bibinfo{year}{2020}\natexlab{}.
\newblock \showarticletitle{Mining API usage scenarios from stack overflow}.
\newblock \bibinfo{journal}{\emph{Information and Software Technology}}
  \bibinfo{volume}{122} (\bibinfo{year}{2020}), \bibinfo{pages}{106277}.
\newblock


\bibitem[Vaswani et~al\mbox{.}(2017)]%
        {vaswani2017attention}
\bibfield{author}{\bibinfo{person}{Ashish Vaswani}, \bibinfo{person}{Noam
  Shazeer}, \bibinfo{person}{Niki Parmar}, \bibinfo{person}{Jakob Uszkoreit},
  \bibinfo{person}{Llion Jones}, \bibinfo{person}{Aidan~N Gomez},
  \bibinfo{person}{{\L}ukasz Kaiser}, {and} \bibinfo{person}{Illia
  Polosukhin}.} \bibinfo{year}{2017}\natexlab{}.
\newblock \showarticletitle{Attention is all you need}. In
  \bibinfo{booktitle}{\emph{Advances in neural information processing
  systems}}. \bibinfo{pages}{5998--6008}.
\newblock


\bibitem[Wang et~al\mbox{.}(2021)]%
        {wang2021automatic}
\bibfield{author}{\bibinfo{person}{Song Wang}, \bibinfo{person}{Nishtha
  Shrestha}, \bibinfo{person}{Abarna~Kucheri Subburaman},
  \bibinfo{person}{Junjie Wang}, \bibinfo{person}{Moshi Wei}, {and}
  \bibinfo{person}{Nachiappan Nagappan}.} \bibinfo{year}{2021}\natexlab{}.
\newblock \showarticletitle{Automatic Unit Test Generation for Machine Learning
  Libraries: How Far Are We?}. In \bibinfo{booktitle}{\emph{2021 IEEE/ACM 43rd
  International Conference on Software Engineering (ICSE)}}. IEEE,
  \bibinfo{pages}{1548--1560}.
\newblock


\bibitem[Xie et~al\mbox{.}(2020)]%
        {xie2020api}
\bibfield{author}{\bibinfo{person}{Wenkai Xie}, \bibinfo{person}{Xin Peng},
  \bibinfo{person}{Mingwei Liu}, \bibinfo{person}{Christoph Treude},
  \bibinfo{person}{Zhenchang Xing}, \bibinfo{person}{Xiaoxin Zhang}, {and}
  \bibinfo{person}{Wenyun Zhao}.} \bibinfo{year}{2020}\natexlab{}.
\newblock \showarticletitle{API method recommendation via explicit matching of
  functionality verb phrases}. In \bibinfo{booktitle}{\emph{Proceedings of the
  28th ACM Joint Meeting on European Software Engineering Conference and
  Symposium on the Foundations of Software Engineering}}.
  \bibinfo{pages}{1015--1026}.
\newblock


\bibitem[Xu et~al\mbox{.}(2020)]%
        {xu2020incorporating}
\bibfield{author}{\bibinfo{person}{Frank~F Xu}, \bibinfo{person}{Zhengbao
  Jiang}, \bibinfo{person}{Pengcheng Yin}, \bibinfo{person}{Bogdan Vasilescu},
  {and} \bibinfo{person}{Graham Neubig}.} \bibinfo{year}{2020}\natexlab{}.
\newblock \showarticletitle{Incorporating External Knowledge through
  Pre-training for Natural Language to Code Generation}. In
  \bibinfo{booktitle}{\emph{Proceedings of the 58th Annual Meeting of the
  Association for Computational Linguistics}}.
\newblock


\bibitem[Yao et~al\mbox{.}(2018)]%
        {yao2018staqc}
\bibfield{author}{\bibinfo{person}{Ziyu Yao}, \bibinfo{person}{Daniel~S Weld},
  \bibinfo{person}{Wei-Peng Chen}, {and} \bibinfo{person}{Huan Sun}.}
  \bibinfo{year}{2018}\natexlab{}.
\newblock \showarticletitle{Staqc: A systematically mined question-code dataset
  from stack overflow}. In \bibinfo{booktitle}{\emph{Proceedings of the 2018
  World Wide Web Conference}}. \bibinfo{pages}{1693--1703}.
\newblock


\bibitem[Yin et~al\mbox{.}(2018)]%
        {yin2018learning}
\bibfield{author}{\bibinfo{person}{Pengcheng Yin}, \bibinfo{person}{Bowen
  Deng}, \bibinfo{person}{Edgar Chen}, \bibinfo{person}{Bogdan Vasilescu},
  {and} \bibinfo{person}{Graham Neubig}.} \bibinfo{year}{2018}\natexlab{}.
\newblock \showarticletitle{Learning to mine aligned code and natural language
  pairs from stack overflow}. In \bibinfo{booktitle}{\emph{2018 IEEE/ACM 15th
  international conference on mining software repositories (MSR)}}. IEEE,
  \bibinfo{pages}{476--486}.
\newblock


\bibitem[Yin and Neubig(2017)]%
        {yin2017syntactic}
\bibfield{author}{\bibinfo{person}{Pengcheng Yin} {and} \bibinfo{person}{Graham
  Neubig}.} \bibinfo{year}{2017}\natexlab{}.
\newblock \showarticletitle{A Syntactic Neural Model for General-Purpose Code
  Generation}. In \bibinfo{booktitle}{\emph{Proceedings of the 55th Annual
  Meeting of the Association for Computational Linguistics (Volume 1: Long
  Papers)}}. \bibinfo{pages}{440--450}.
\newblock


\bibitem[Yin and Neubig(2018)]%
        {yin2018tranx}
\bibfield{author}{\bibinfo{person}{Pengcheng Yin} {and} \bibinfo{person}{Graham
  Neubig}.} \bibinfo{year}{2018}\natexlab{}.
\newblock \showarticletitle{TRANX: A Transition-based Neural Abstract Syntax
  Parser for Semantic Parsing and Code Generation}. In
  \bibinfo{booktitle}{\emph{Proceedings of the 2018 Conference on Empirical
  Methods in Natural Language Processing: System Demonstrations}}.
  \bibinfo{pages}{7--12}.
\newblock


\bibitem[Zan et~al\mbox{.}(2022)]%
        {zan2022cert}
\bibfield{author}{\bibinfo{person}{Daoguang Zan}, \bibinfo{person}{Bei Chen},
  \bibinfo{person}{Dejian Yang}, \bibinfo{person}{Zeqi Lin},
  \bibinfo{person}{Minsu Kim}, \bibinfo{person}{Bei Guan},
  \bibinfo{person}{Yongji Wang}, \bibinfo{person}{Weizhu Chen}, {and}
  \bibinfo{person}{Jian-Guang Lou}.} \bibinfo{year}{2022}\natexlab{}.
\newblock \showarticletitle{CERT: Continual Pre-Training on Sketches for
  Library-Oriented Code Generation}.
\newblock \bibinfo{journal}{\emph{arXiv preprint arXiv:2206.06888}}
  (\bibinfo{year}{2022}).
\newblock


\bibitem[Zhang et~al\mbox{.}(2020)]%
        {zhang2020machine}
\bibfield{author}{\bibinfo{person}{Jie~M Zhang}, \bibinfo{person}{Mark Harman},
  \bibinfo{person}{Lei Ma}, {and} \bibinfo{person}{Yang Liu}.}
  \bibinfo{year}{2020}\natexlab{}.
\newblock \showarticletitle{Machine learning testing: Survey, landscapes and
  horizons}.
\newblock \bibinfo{journal}{\emph{IEEE Transactions on Software Engineering}}
  (\bibinfo{year}{2020}).
\newblock


\end{thebibliography}










\end{document}